\documentclass[aps,prd,showpacs,amssymb,superscriptaddress, notitlepage,floats,floatfix,nofootinbib, twocolumn,groupedaddress, longbibliography]{revtex4-2}

\usepackage{dcolumn}% Align table columns on decimal point
\usepackage{bm}% bold math
\usepackage{amsmath}	% Advanced maths commands
\usepackage{amssymb}
\usepackage[utf8]{inputenc}
\usepackage{graphicx}
\usepackage{verbatim}
\usepackage{spverbatim}
\usepackage{float}
\usepackage{placeins}
\usepackage[dvipsnames]{xcolor}
\usepackage{soul}
\usepackage{xfrac}
\usepackage{changepage}
\usepackage{mathtools, wasysym}
\usepackage{relsize}
\usepackage{amsmath,tabularx,booktabs}
\usepackage{lipsum}
\usepackage[normalem]{ulem}

\newcolumntype{C}{>{\Centering\hspace{0pt}}X}

    \newcommand{\gtsima}{$\; \buildrel > \over \sim \;$}
\newcommand{\ltsima}{$\; \buildrel < \over \sim \;$}
\newcommand{\prosima}{$\; \buildrel \propto \over \sim \;$}
\newcommand{\gsim}{\lower.5ex\hbox{\consistegtsima}}
\newcommand{\lsim}{\lower.5ex\hbox{\ltsima}}
\newcommand{\simgt}{\lower.5ex\hbox{\gtsima}}
\newcommand{\simlt}{\lower.5ex\hbox{\ltsima}}
\newcommand{\simpr}{\lower.5ex\hbox{\prosima}}
\newcommand{\sgrA}{Sgr\,A$^\star$}
\newcommand{\mpbh}{m_{\textrm{\scalebox{.75}{PBH}}}}
\newcommand{\lw}{\scalebox{.55}}
\newcommand{\txt}{\scalebox{.75}}
\newcommand{\comm}[1]{}

\graphicspath{ {armoniche/} }
\usepackage{array}
\newcolumntype{x}[1]{>{\centering\let\newline\\\arraybackslash\hspace{0pt}}p{#1}}
\newcolumntype{y}[1]{>{\raggedright\let\newline\\\arraybackslash\hspace{0pt}}p{#1}}

\newcommand{\GW}{\mathrm{GW}}

\begin{document}
\title{Gravitational waves from an eccentric population of\\ primordial black holes orbiting \sgrA 
}
\author{Stefano Bondani$^1$}
\author{Matteo Bonetti$^{3,2}$}
\author{Luca Broggi$^{3,2}$}
\author{Francesco Haardt$^{1,2}$}
\author{Alberto Sesana$^{3,2}$}
\author{Massimo Dotti$^{3,2}$}
\affiliation{$^1$DiSAT, Universit\`a degli Studi dell'Insubria, Via Valleggio 11, 22100 Como, Italy}
\affiliation{$^2$INFN, Sezione Milano-Bicocca, Piazza della Scienza 3, Milano 20126, Italy}
\affiliation{$^3$Dipartimento di Fisica ``G. Occhialini'', Universit\`a degli Studi di Milano-Bicocca, Piazza della Scienza 3, Milano 20126, Italy}
\date{\today}

\begin{abstract}
\noindent Primordial black holes (PBH), supposedly formed in the very early Universe, have been proposed as a possible viable dark matter candidate. In this work we characterize the expected gravitational wave (GW) losses from a population of PBHs orbiting \sgrA, the super-massive black hole at the Galactic center (GC), and assess the signal detectability by the planned space-borne interferometer LISA and by the proposed next generation space-borne interferometer $\mu$Ares. Assuming that PBHs indeed form the entire diffuse mass allowed to reside within the orbit of the S2 star, we compute an upper limit to the expected GW signal both from resolved and non-resolved sources, under the further assumptions of monochromatic mass function and thermally distributed eccentricities. By comparing with our previous work where PBHs on circular orbits were assumed, we show for 1\,M$_{\odot}$ PBHs how the GW signal from high harmonics over a 10 year data stream increases by a factor of six the chances of LISA detectability, from the $\approx 10\%$ of the circular case, to $\approx 60\%$, whereas multiple sources can be identified in 20\% of our mock populations. The background signal, made by summing up all non resolved sources, should be certainly detectable thanks to the PBHs with higher eccentricity evolving under two body relaxation. In the case of $\mu$Ares, because of its improved sensitivity in the $\mu$Hz band, one third of the entire population of PBHs orbiting \sgrA~would be resolved. The background noise from the remaining non resolved sources should be detectable as well. Finally we present the results for different PBH masses.
\end{abstract}
\maketitle
\section{\label{sec_intro}introduction}
The search for a solution to the dark matter problem \cite{reviewdm, directdm1, directdm2, kamioka} has led primordial black holes (PBHs) \cite{zeldovic, zeldovic66, hawkingPBH, carrhawking, hawkingbousso} to be considered as a serious candidate for several years now \cite{hawkins}, especially in light of the as-of-today inconclusive results of the many experiments aimed at a direct detection of a dark matter particle \cite{XENONnT, PandaX}. If PBHs do exist and indeed constitute a fraction of the dark matter in the Universe, it is argued that they should clump at the center of galaxies, and supposedly orbit the supermassive black hole at their center \cite{wang_2020, bondani, freese, freeseLISA, maselli}, with a mass function and abundance given by, e.g., \cite{carr2021}. Given the PBHs compact nature, their motion around the central supermassive black hole should be accompanied by loud gravitational wave (GWs) emission, whose detection, or lack thereof, could put constraints on the very existence of PBHs and on their relevance as a dark matter candidate. 

Closely following the approach in \cite{wang_2020}, we characterized in \cite{bondani} the GW signal in the LISA and $\mu$Ares interferometers expected in the most relevant case of \sgrA, the supermassive black hole at the center of the Milky Way, by allowing for a total possible diffuse mass of $4\times10^3$\,M$_{\odot}$ inside a $10^{-3}$\,pc (i.e., within the pericenter of star S2, see \cite{gravity21}), entirely comprised of PBHs, and assuming monochromatic mass functions for the resident PBH population on circular orbits. In particular, for the most physically relevant case of 1\,$M_{\odot}$ PBHs, a 10-years long LISA mission was found to have a 10\% chance of detection of one single PBH, while $\mu$Ares was expected to be able to resolve more than 140 of them. The GW background was shown to have a signal to noise ratio (SNR) largely undetectable by LISA, while $\mu$Ares would be sensitive enough to detect it with SNR well over 100.

In this work we expand on our previous results. While maintaining consistency in all other model parameters and assumptions in order to facilitate a meaningful comparison, we consider the more physically sound scenario of PBHs on eccentric orbits, characterized here by a thermal distribution in eccentricity, i.e., $p(e)\propto e$, as predicted by  general evolutionary arguments, see e.g. \cite{thermal_jeans, thermal_Geller, thermal_bormio}.
The effect of $e$ is expected to be twofold. First, since the GW timescale is a strong function of $e$, PBHs will fall towards \sgrA~due to GW losses starting from different distances and at different pace. Second, eccentric binaries emit GWs at harmonics higher than the orbital frequency, thus shifting most of GW radiation at higher frequencies.

In the present work we will assess the expected signal from all resolvable sources and the residual contribution to GW background noise\footnote{Consistently to our previous work, we assume here observation duty cycles of 10 years for both facilities.}, adopting the spectral sensitivity curves of forthcoming/proposed space-based observatories such as LISA \cite{2017arXiv170200786A} and $\mu$Ares \cite{muares}.

We must stress, however, that by allowing the entire diffuse mass around \sgrA~to be in PBHs, our results should be taken as an upper bound on the foreseeable detectability of these objects with future interferometers. Furthermore, the assumption that a spiked dark matter density profile is actually present at the GC is an unproved and rather debated one \cite{blok, perspective, spiral, revisited}, which implies a possibly overestimated dark matter density at the GC, when considering a local density of $\simeq 0.5$\,GeV\,cm$^{-3}$ \cite{read2014}. 

The paper is structured as follows. In Section\,\ref{sec_eccdensity} we derive the number density distribution of PBHs in the GW-dominated region. In Section\,\ref{sec_eccGW} we review the theoretical background behind the generation of the GW signal from a binary of given eccentricity. In Section\,\ref{sec_eccresults} we present the results and quantify the detectability of the computed GW resolved and background signal. Finally, Section\,\ref{sec_concl} is dedicated to the discussion of our results and to conclusive remarks. 

\section{\label{sec_eccdensity}Evolution of PBH density distribution}
In continuity with \cite{wang_2020, bondani} we assume a population of PBHs characterized by a monochromatic mass function, $\mpbh=1\,M_{\odot}$ \cite{prefmass}. PBHs are radially distributed such as, at large distances, far  from the GW-dominated regime, the density profile (referred to as \textit{unperturbed}) is driven by two-body relaxation. The timescale over which two-body relaxation acts, is obtained from the relaxation timescale for circular orbits~\cite{binney}:
%%%%%%%%%%%%%%%%%
\begin{equation} 
\tau_{\txt{rel,c}} = \frac{1.8\times10^{10}\,\textrm{yr}}{\log(M_{\txt{MBH}}/\mpbh)} \frac{1M_{\odot}}{\mpbh}\frac{10^{3} M_{\odot}\,\textrm{pc}^{-3}}{\rho}\left(\frac{\sigma}{10 \,{\rm km\,s^{-1}}}\right)^{3}
\label{eq:t2br}
\end{equation}
where $\rho$ is the density distribution of the PBH population, in our case the very same spiked Navarro-Frenk-White (NFW, \cite{nfw}) profile \cite{gondolo} adopted in \cite{wang_2020, bondani}, and $\sigma$ is the Keplerian velocity, both functions of the semi-major axis $a$; the mass of the supermassive black hole, in our case \sgrA, is $M_{\txt{MBH}}=4.3\times10^6\,M_{\odot}$ \cite{gravity21}. From Eq.\,(\ref{eq:t2br}), the two-body relaxation timescale for eccentric orbits is readily obtained as 
\begin{equation}
\tau_{\txt{rel,e}}=\tau_{\txt{rel,c}}\times(1-e).
\end{equation}
If now we set $a_0$ and $e_0$ as the initial values of semi-major axis and eccentricity, the timescale for GW infall is given by \cite{Maggiore2007}:
\begin{equation}
    \tau_{\lw{GW}}(e_0,a_0)=\tau_0(a_0)F(e_0),
\end{equation}
where
\begin{equation}
    \tau_0(a_0)=\frac{5}{256}\frac{c^5a_0^4}{G^3m^2\mu}
\end{equation}
is the timescale for GW coalescence for circular orbits for a binary of total mass $m$ and reduced mass $\mu$, and the term containing the eccentricity dependency is
\begin{equation}
    \label{eq:F0}
    F(e_0)=\frac{48}{19}\frac{1}{g^4(e_0)}\int_0^{e_0}de\frac{g^4(e)(1-e^2)^{5/2}}{e(1+\frac{121}{304}e^2)}, 
\end{equation}
in which the function of eccentricity $g(e)$ appearing in Eq.\,(\ref{eq:F0}) is given by:
\begin{equation}
    \label{eq:ge}
    g(e)=\frac{e^{12/19}}{1-e^2}\Bigl(1+\frac{121}{304}e^2\Bigr)^{870/2299}.
\end{equation}

\begin{figure}
{\includegraphics[scale=0.45]{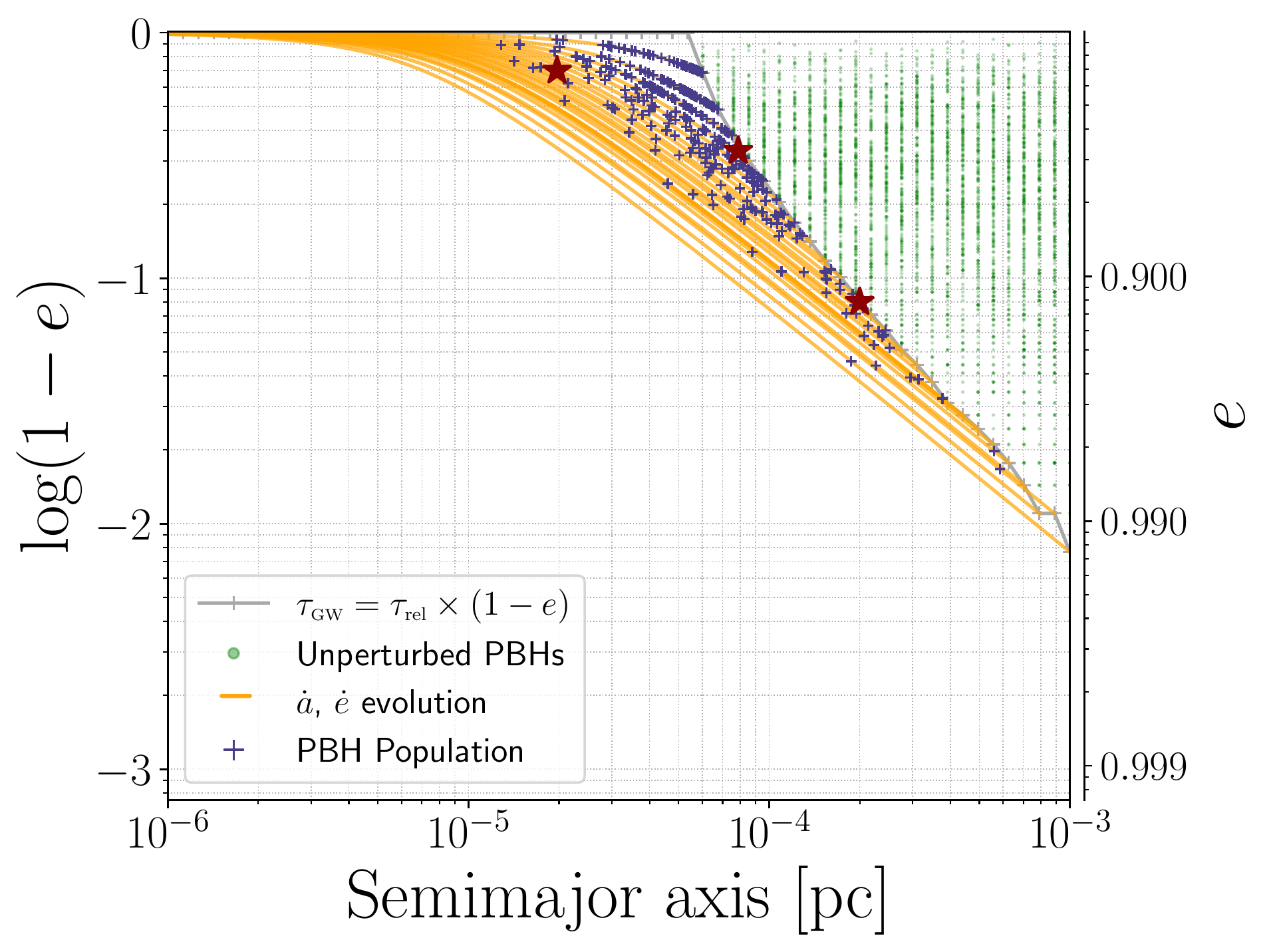}
\vspace{-.5cm}
\caption{\label{fig:popolazione}PBH population, unperturbed (green dots) and inside the GW regime (purple markers), randomly sampled over their own time to merger $\tau_{\lw{GW}}$. %which can be read vertically on the red crosses and axis.
The grey demarcation line indicates where $\tau_{\lw{GW}}=\tau_{\txt{rel}}\times(1-e)$, while the orange evolution tracks are obtained by solving the coupled differential equations for $\dot{a}$ and $\dot{e}$. The three different red markers refer to representative examples of low, average and high initial eccentricity, as detailed in the text.}
}
\end{figure}

For each semi-major axis $a$, we then define $e_{\lw{GW}}(a)$ as the eccentricity such that
%$\tau_{\lw{GW}}$ and $\tau_{\txt{rel}}\times(1-e)$ are equal.
\begin{equation}
    \tau_{\lw{GW}}\left(a, e_{\lw{GW}}\right) = \tau_{\txt{rel,c}}(a)\times\left(1 - e_{\lw{GW}}\right)\equiv\tau_{\txt{rel,e}}.
\label{eq:eGW}
\end{equation}
This limiting eccentricity effectively divides the parameter space semi-major axis-circularity (i.e., $[a,1-e]$) into two regions, as shown by the grey line in Fig.\,\ref{fig:popolazione}. For eccentricities larger than $e_{\lw{GW}}$ the orbital evolution is driven by GW emission. Time dependent $a(t)$ and $e(t)$ can be obtained by solving  the following coupled differential equations (orange trajectories in Fig.\,\ref{fig:popolazione}):
\begin{equation}
    \dot{a} = -\frac{64}{5}\frac{G^3\mu m^2}{c^5a^3}\frac{1}{(1-e^2)^{7/2}}\Bigl(1+\frac{73}{24}e^2+\frac{37}{96}e^4\Bigr)
\label{eq:adot}
\end{equation}
and
\begin{equation}
    \dot{e} = -\frac{304}{15}\frac{G^3\mu m^2}{c^5a^4}\frac{e}{(1-e^2)^{5/2}}\Bigl(1+\frac{121}{304}e^2\Bigr). 
\label{eq:edot}
\end{equation}

%%%%%%%%%%%%%%%%%%%%%%%%%%%%
%%%%%%%%%%%%%%%%%%%%%%%%%%%%%%%%%

In Fig.\,\ref{fig:popolazione}, on the right side of the grey dividing line in the $[a, 1-e]$ plane, we show as green dots an example of a Monte Carlo-sampled unperturbed population of PBHs, distributed on a grid of semi-major axis according to the spiked NFW density profile; subsequently, in each bin the eccentricity has been drawn Monte Carlo from a thermal distribution $dp/de=2e$. The total PBH population is normalized to 4,000 $\mpbh=1\,M_{\odot}$ PBHs within a $10^{-3}$\,pc distance from \sgrA~\cite{gravity21}. On the left side of the same dividing line, the purple markers indicate instead the positions of those PBHs whose evolution is dominated by GW emission. These positions along the evolutionary tracks, here depicted as orange lines, have been Monte Carlo-sampled over their GW merging time $\tau_{\lw{GW}}$. The derivation of the rate at which PBHs cross the dividing line between two body relaxation and GW-dominated regimes, and hence the PBH position in the $[a,1-e]$ plane, will be detailed in the next subsection.
\subsection{Crossing rate}
\begin{figure}
\vspace*{-.6cm}
\hspace*{-.4cm}
{\includegraphics[scale=0.55]{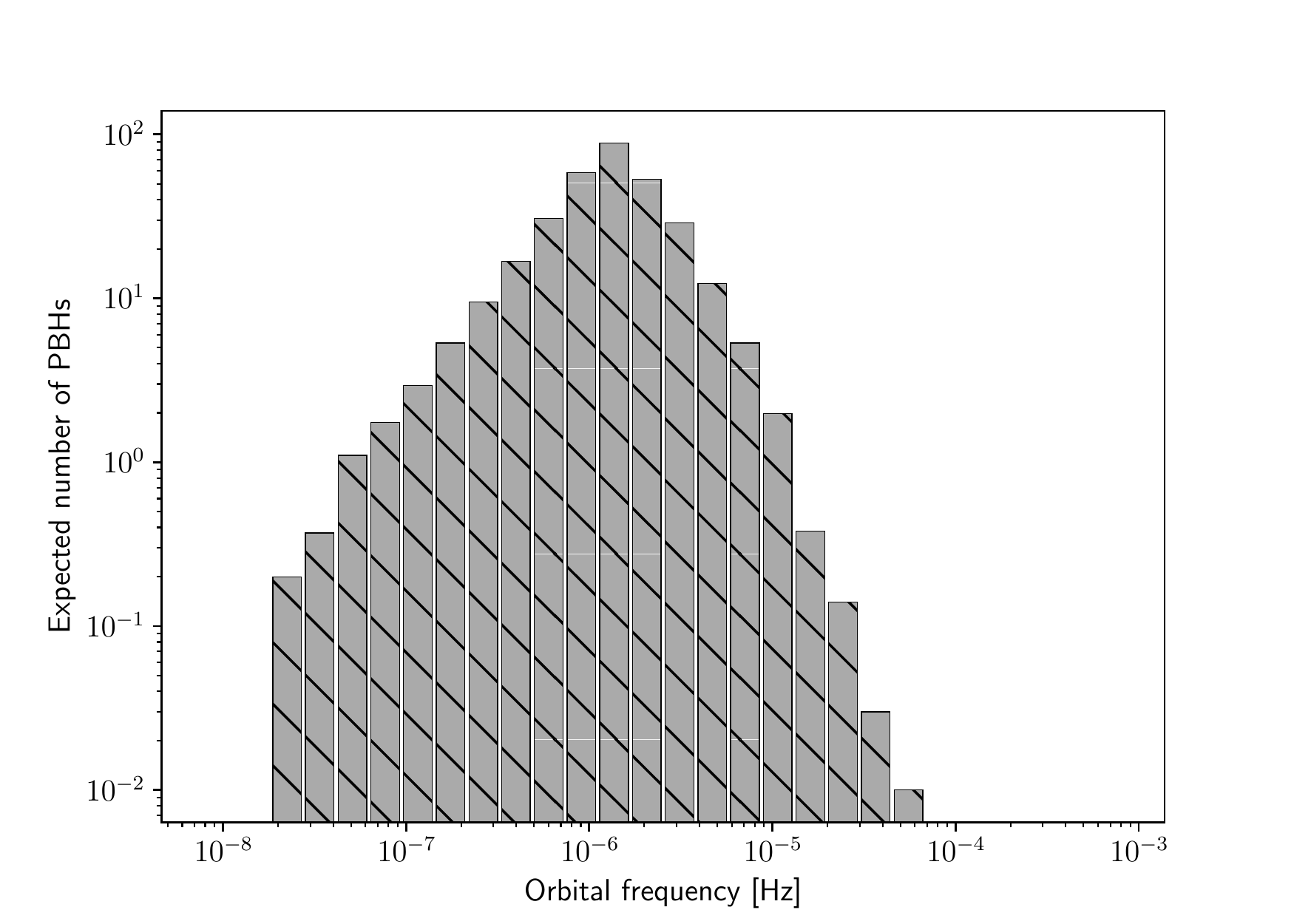}}
\vspace*{-.6cm}\caption{Expected orbital frequency distribution of sources in the GW domain.}
\label{fig:freq_prob_100MC}
\end{figure}
The actual number of PBHs evolving because of GWs emission is obtained by first modelling the process of two-body relaxation acting at larger scales from the central supermassive BH, hence providing the infall rate of objects. To this goal we need to assume an underlying density profile which, as in~\cite{bondani}, follows a spiked NFW~\cite{gondolo}. We then impose the Cohn-Kalsrud condition~\cite{cohn, stone} to estimate the inward flux.
\begin{figure}[!b]
\vspace{-.05cm}
\hspace{-.3cm}
{\includegraphics[scale=0.51]{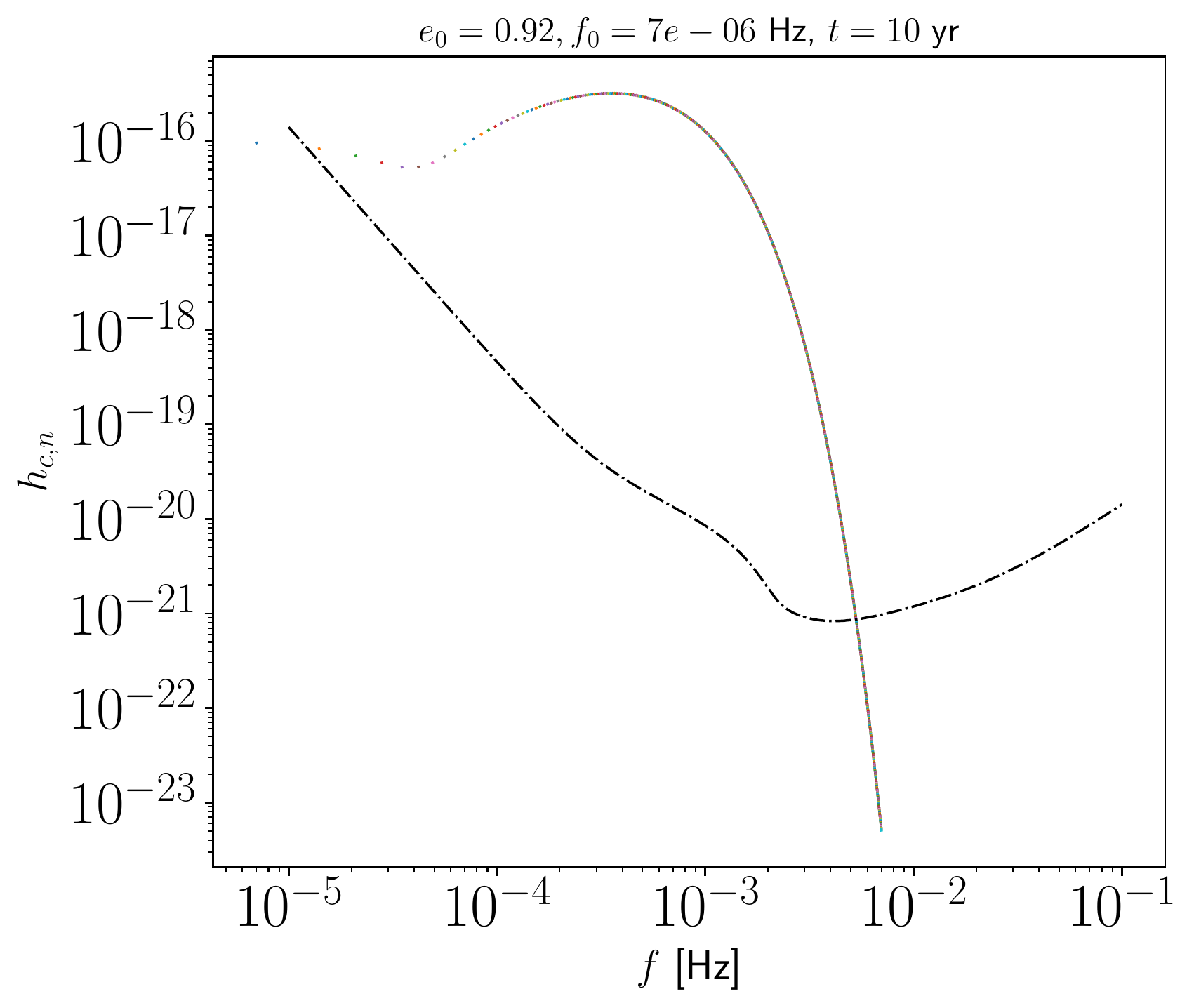}}
\vspace{-.7cm}
\caption[Example harmonics]{\label{fig:hcharm}Example of characteristic strains at every (of the first 1,000) harmonic referred to a 1\,M$_{\odot}$ PBH of initial eccentricity $e_0=0.92$, initial orbital frequency $f_0=7\times10^{-6}$\,Hz and total
observation time $t=$ 10 years with LISA (sensitivity curve plotted for reference in black). The colored segments show how the frequency shift over 10 years of each harmonic is very limited.}
\end{figure}

In more detail, Cohn and Kulsrud \cite{cohn} provided a framework to describe the stochastic evolution of stellar objects around a central massive black hole. In particular, they adopted the orbit-averaged Fokker-Planck equation \cite{fokkerplanck} as an evolution equation for the distribution function of these objects. In the formalism, an object is considered to be captured (and thus does not contribute to relaxation anymore) when its pericenter is smaller than a critical radius $r_c$, usually referred to as the loss-cone radius, modeled on the physics of the capture process, e.g.~tidal disruption events or capture of compact objects. The criterion can be expressed in terms of orbital parameters, so that a particle with energy $E$ is considered to be captured when relaxation drives its eccentricity above a critical value $e_c(E)$. Since relaxation in angular momentum for these systems is more efficient than relaxation in energy (see also \citep{merritt_textbook, stone}), \cite{cohn} showed that over timescales of order of $\tau_{\txt{rel}}$ (from now on indicating $\tau_{\txt{rel},c}$ as simply $\tau_{\txt{rel}}$) the distribution function assumes a quasi-stable profile in eccentricity, depending upon $e_{c}$ and on the local relaxation rate $1/\tau_{\txt{rel}}$. From the resulting equilibrium distribution it is then possible to compute the rate of objects captured by the central black hole 
(see Eqs.\,(15-17) in \cite{stone}):
\begin{equation}
    \frac{dN}{dEdt}= \frac{\bar{N}}{\tau_{\txt{rel}}(a) \left[\log \frac{1}{1-e_{\lw{c}}^2} - (1-\alpha)\,e_{\lw{c}}^2\right]}
    \label{eq:dNdEdt_},
\end{equation}
where $\bar{N}$ is the number of objects per unit energy, such that its integral over $E$ provides the total number of objects and
\begin{equation}
    \alpha \simeq \sqrt{\frac{P(a)}{\tau_\mathrm{rel}(a)}}
\end{equation}
is a parameter expressing the efficiency of diffusion at a given orbit with radial period $P(a)$. Since $P(a)$ is small very close to the massive BH, we consider $\alpha \simeq 0$.
In the standard treatment of the capture of compact objects, it is customary (e.g.,~\cite{broggi_extreme_2022}) to set the critical eccentricity so that $r_c= 8 \, G M_{\txt{MBH}} /c^2$, retaining diffusion up to general relativistic scales ~\cite{bar-or_steady_2016}.
In our model, on the other hand, particles with a given semi-major axis enter the phase of GW-driven evolution with $e_{\lw{GW}}$ implicitly given by Eq.\,(\ref{eq:eGW}) and are no more subject (nor contribute) to two-body relaxation; therefore in this process $e_c=e_\GW(a)$.

We now integrate the above Eq.\,(\ref{eq:dNdEdt_}) over discrete energy bins $\Delta E_i=E_{i+1}-E_i$ (corresponding to discrete bins in separation since $a=GM_{\textrm{MBH}}/2E$).
\begin{figure*}[!t]
    \begin{minipage}[l]{.670\columnwidth}
        \centering
        \includegraphics[scale=0.275]{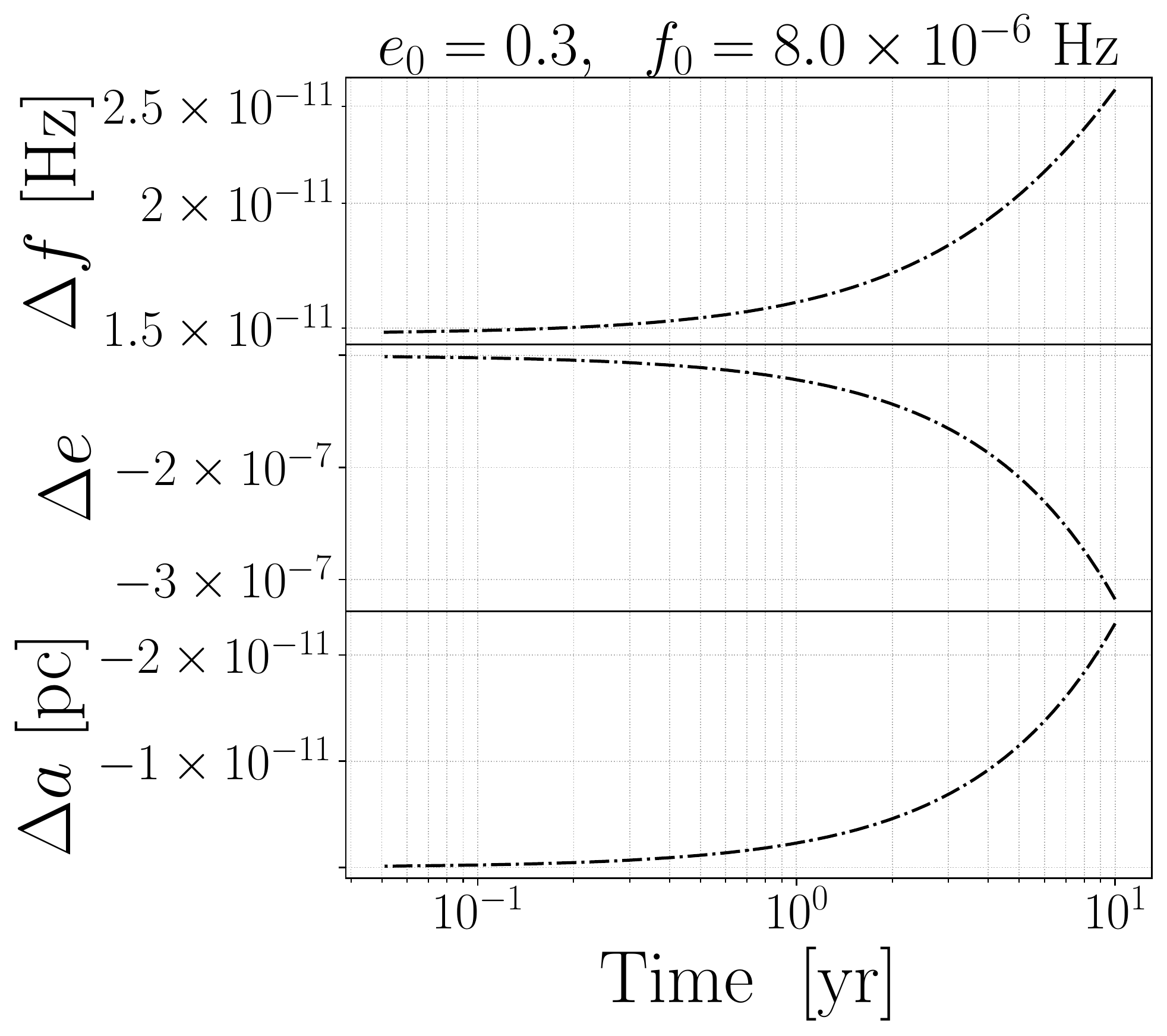}
        \label{fig:cfrA}
    \end{minipage}
    % \hfill{}\hfill{}
    \begin{minipage}[l]{0.670\columnwidth}
        \centering
        \includegraphics[scale=0.275]{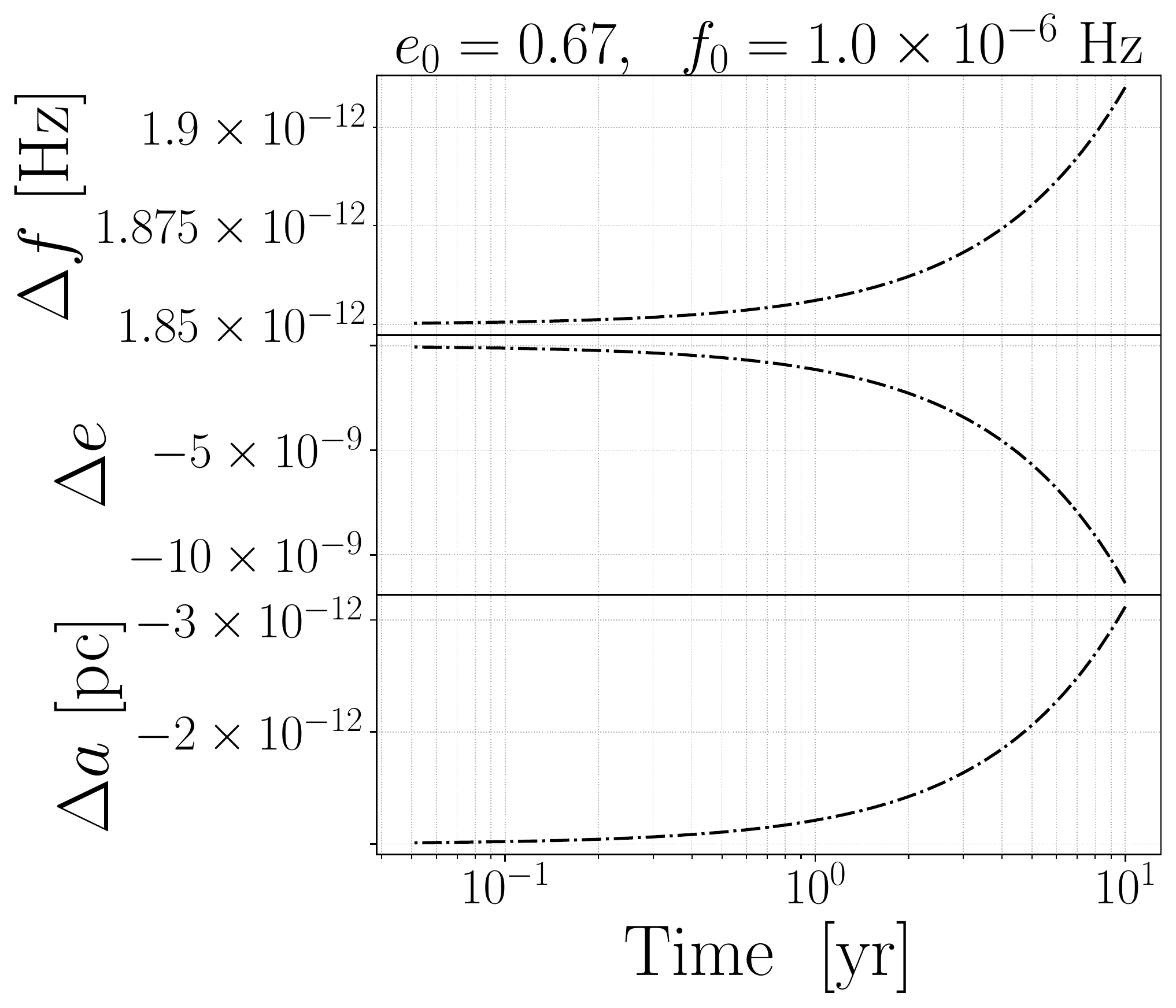}
        \label{fig:cfrB}
    \end{minipage}
    % \hfill{}
    \begin{minipage}[l]{0.67\columnwidth}
        \centering
        \includegraphics[scale=0.275]{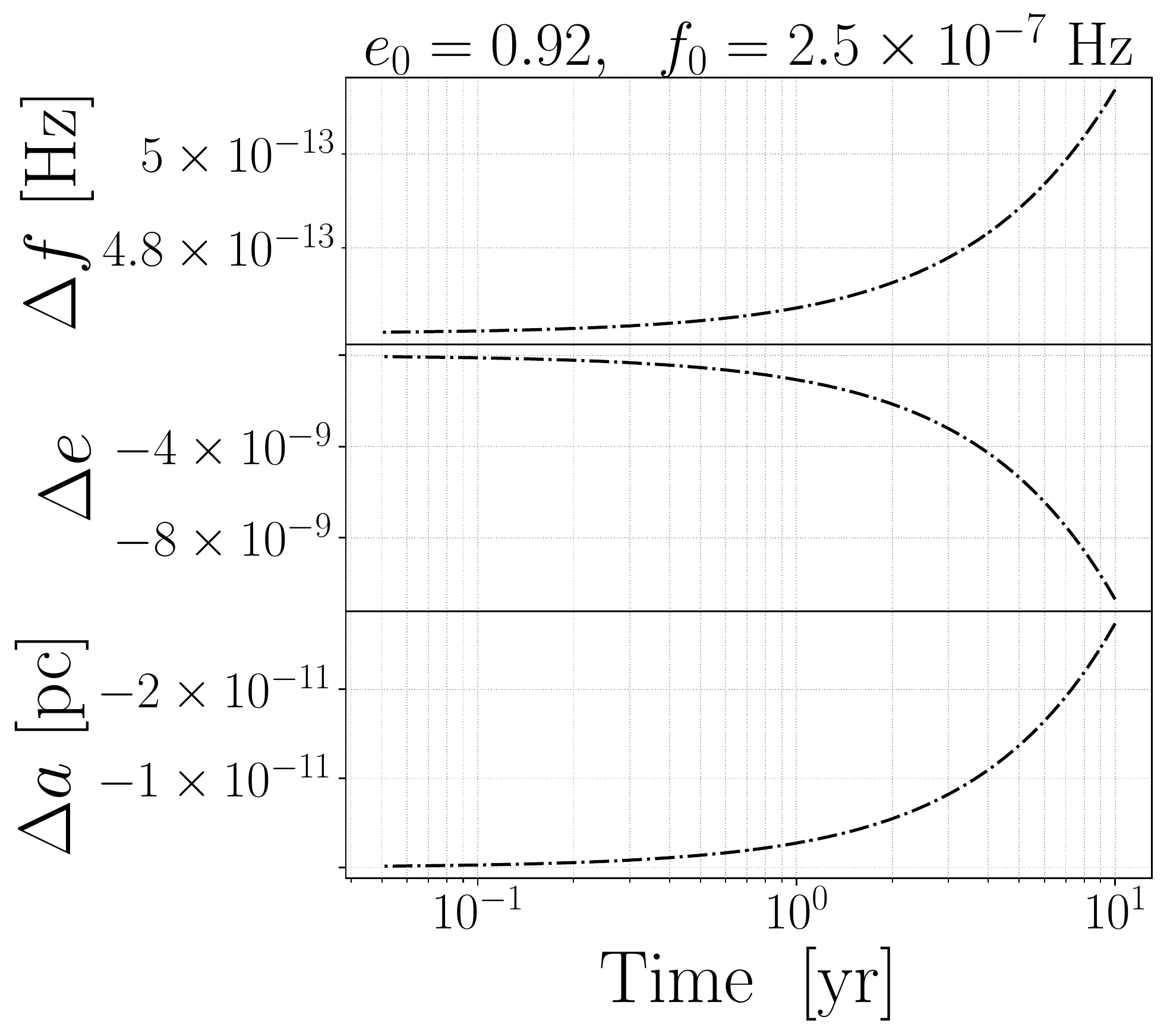}
        \label{fig:cfrC}
    \end{minipage}
    % \vspace*{-5mm}
    \caption{\label{fig:cfr_evo}Evolution of frequency, eccentricity and semi-major axis over 10 years for three different initial values of eccentricity $e_0$. The initial values of orbital frequency from left to right are $8\times10^{-6}$, $1\times10^{-6}$ and $2.5\times10^{-7}$\,Hz, respectively, in agreement with a generic PBH population as depicted in Fig.\,\ref{fig:popolazione}, where they are represented as the three red markers.  }
\end{figure*}
The rate of PBHs crossing into the GW-dominated regime at separation comprised within $a_i$ and $a_{i+1}$ from \sgrA~is then:
\begin{equation}
    \frac{d N_i}{dt} =\frac{\bar N\Delta E_i}{\tau_{\txt{rel}}(a) \left[\log \frac{1}{1-e_{\lw{c}}^2} - e_{\lw{c}}^2\right]}.
\end{equation}\\

Finally, the actual number $N_{\lw{GW}}$ of PBHs found at any time along a specific track given by Eqs.\,(\ref{eq:adot}) and (\ref{eq:edot}) is given by:
\begin{equation}
\label{eq:object_in_GW}
    N_{\lw{GW}}=\frac{dN_i}{dt}\times\tau_{\lw{GW}}(a_i) = \bar N\Delta E_i \, \frac{1-e_c}{\log \frac{1}{1-e_c^2} - e_c^2}, 
\end{equation}
since, at the critical eccentricity, $\tau_{\lw{GW}}=\tau_{\txt{rel,c}}\times(1-e_c)$ by definition.
Eq.\,\eqref{eq:object_in_GW} implies that the number of objects in the GW-dominated regime diverges like $e_c^{-4}$ as $e_c \to 0$. In this work there was no need to address directly this divergence, since the first bin of the distribution of PBHs we consider is centered at $a = 10^{-6}$\,pc, where the critical eccentricity is $e_{c, \txt{min}} \simeq 0.3$ and thus, far from the pole. However, it is worth mentioning that the divergence in Eq.\,\eqref{eq:object_in_GW} disappears in a more accurate model where one accounts for the facts that (\textit{i}) the number of particles in the diffusive regime vanishes for $a\to a_c$ such that $e_c=0$, and that (\textit{ii}) $\alpha$ in Eq.\,\eqref{eq:dNdEdt_} is finite (and not zero). A simple assumption could be that, at a given $a$, only particles  with eccentricity smaller than the critical value $e_c$ participate to diffusion, so that in a thermal distribution $\bar{N} = N(a)\int_0^{e_c} de \, e = N(a)\, e_c^2/2$, and keeping a finite $\alpha$ one gets
\begin{equation}
    N_{\lw{GW}} \to \frac{N(a)\Delta E}{2} \, \frac{1}{\alpha} \qquad e_c \to 0 \, .
\end{equation}

Typically, starting from 4,000 PBHs lying outside the GW-dominated regime but within a 10$^{-3}$\,pc distance from \sgrA~(the green dots in Fig.\,\ref{fig:popolazione}), approximately 300 will migrate inside the GW-dominated region over a timescale $\tau_{\lw{GW}}(a_i)$ (the purple markers in Fig.\,\ref{fig:popolazione}). Their corresponding orbital frequency distribution is plotted in Fig.\,\ref{fig:freq_prob_100MC}.
\begin{figure}[H]
\hspace*{-0.3cm}
{\includegraphics[scale=0.38]{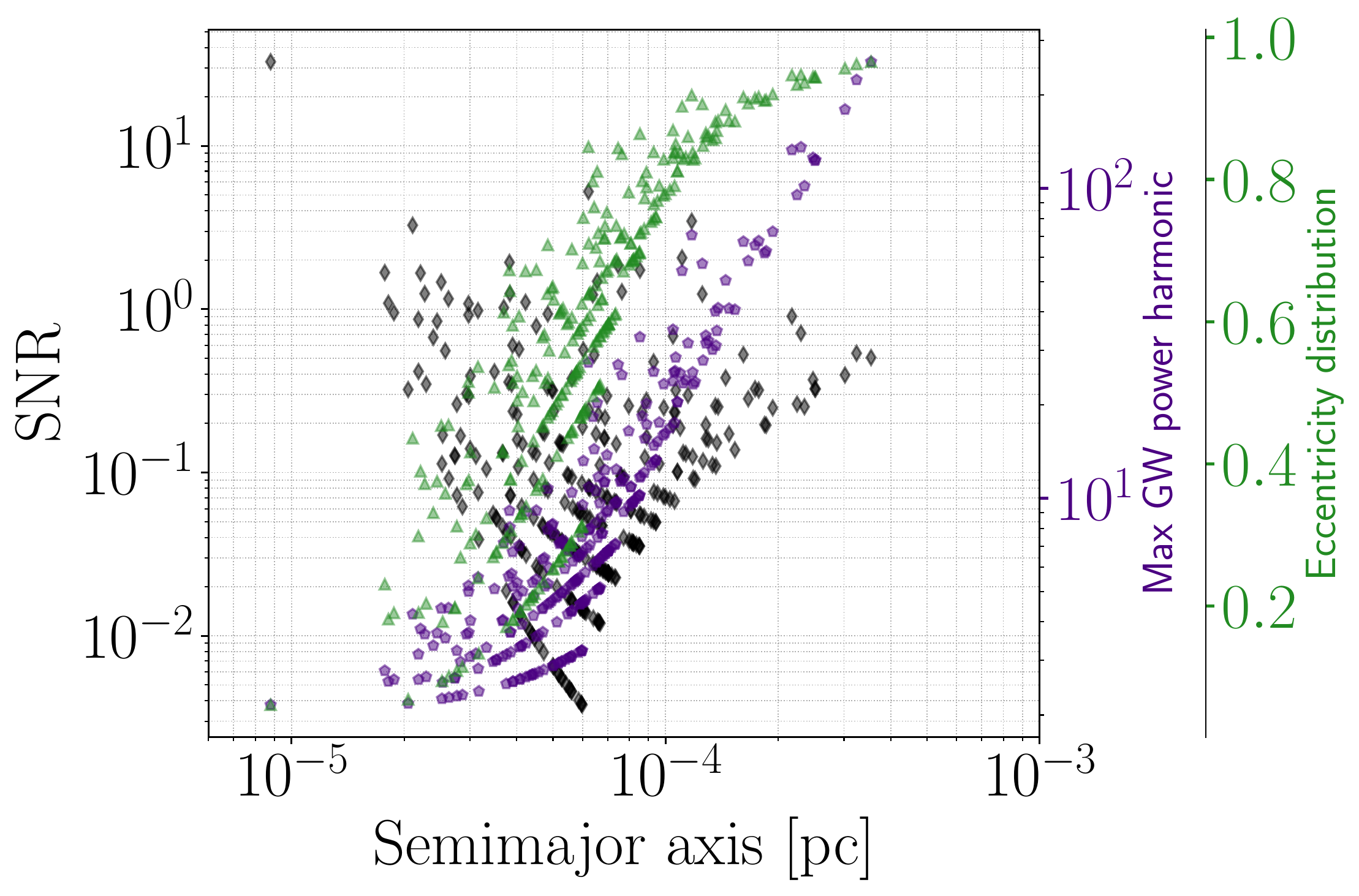}}
\caption{Distributions of SNR (black diamonds), eccentricity (green triangles) and harmonic of maximum GW power emission (purple stars) over semi-major axis of all PBHs evolving by GW emission for a random 1\,M$_{\odot}$~PBH population. To each PBH correspond three different color markers in this plot, vertically aligned with their corresponding semi-major axis.} 
\label{fig:res_SNR}
\end{figure} 
\FloatBarrier
%%%%%%%%%%%%%%%%%%%%%%%%%%%%%%%%%%%%%%%%%%%%%%
\section{Gravitational wave signal}
\label{sec_eccGW}
{The total strain of the emitted GW signal can be characterised at the leading order as a sum over integer harmonics of the orbital Keplerian frequency
\begin{equation}
    h_{c}=\sum_{n=1}^{\infty}h_{c,n},
    \label{eq:hc}
\end{equation}
where the characteristic strain of the $n$-th harmonic is given by \cite{barackcutler2004, bonetti}}:
\begin{equation}
    h_{c,n}=\frac{1}{\pi d}\Bigl(\frac{2G\dot{E_n}}{c^3\dot{f_n}}\Bigr)^{1/2}.
    \label{eq:hcn}
\end{equation}
\begin{figure*}[!ht]
    \begin{minipage}[l]{.99\columnwidth}
        \centering
        \vspace*{-.77cm}
        \includegraphics[scale=0.52]{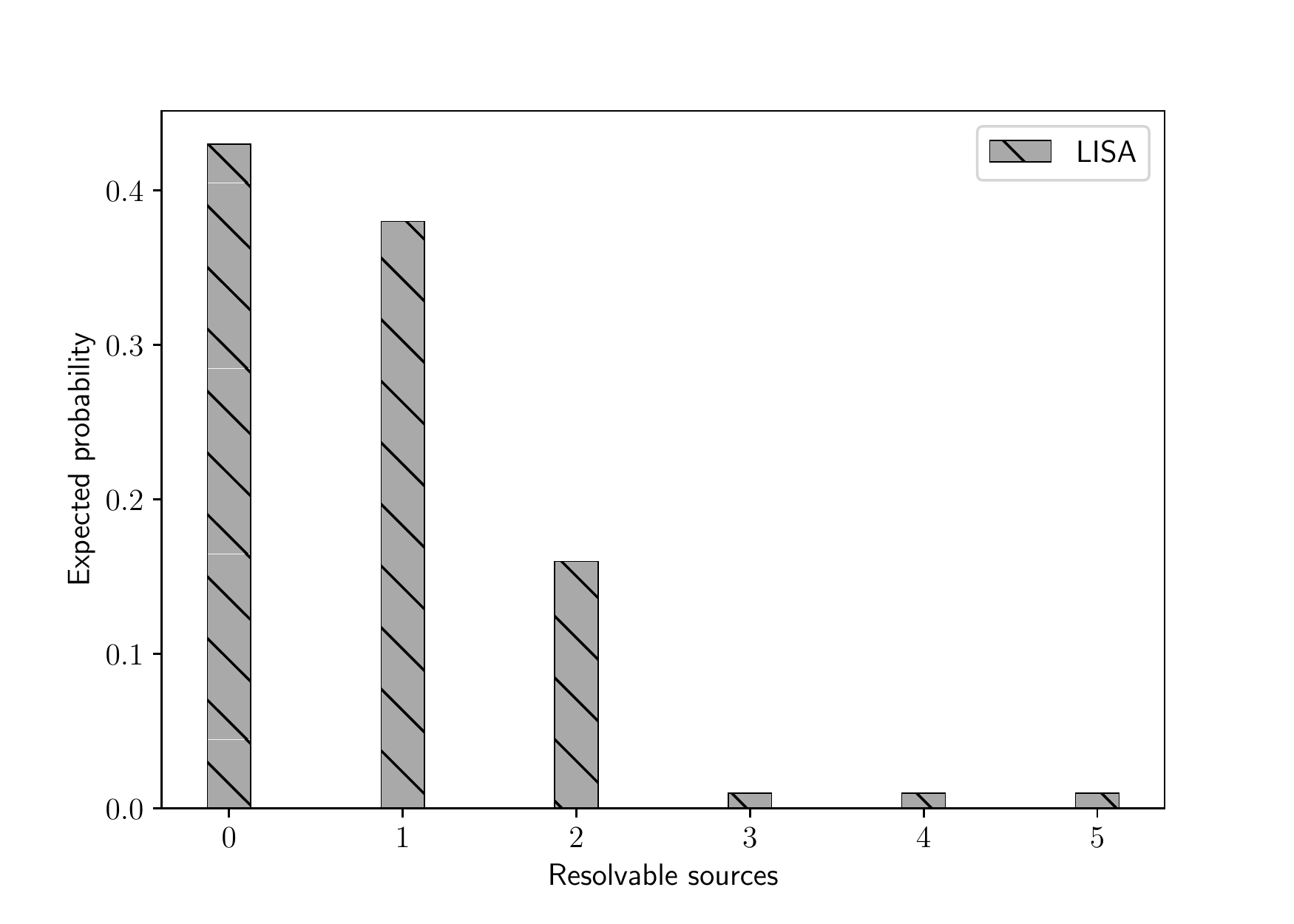}\label{fig:res_LISA_number_100MC}
    \end{minipage}
    % \hfill{}\hfill{}
    \begin{minipage}[l]{0.99\columnwidth}
        \centering
        \vspace{-.5cm}
        \includegraphics[scale=0.51]{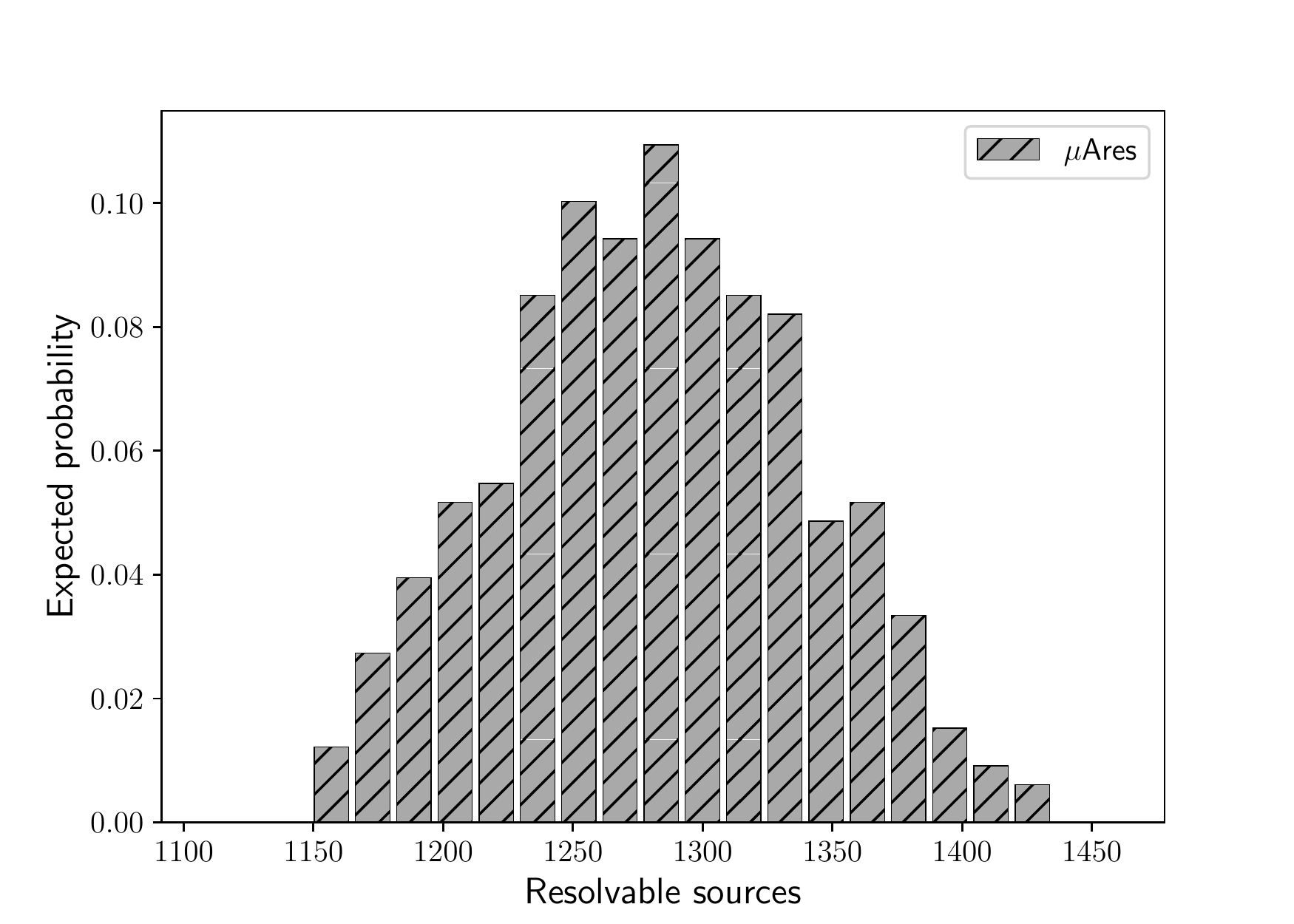}
    \label{fig:res_ARES_number_100MC}
    \end{minipage}\\
    % \hfill{}
    \begin{minipage}[l]{0.99\columnwidth}
        \centering
        \vspace{-.41cm}
        \includegraphics[scale=0.52]{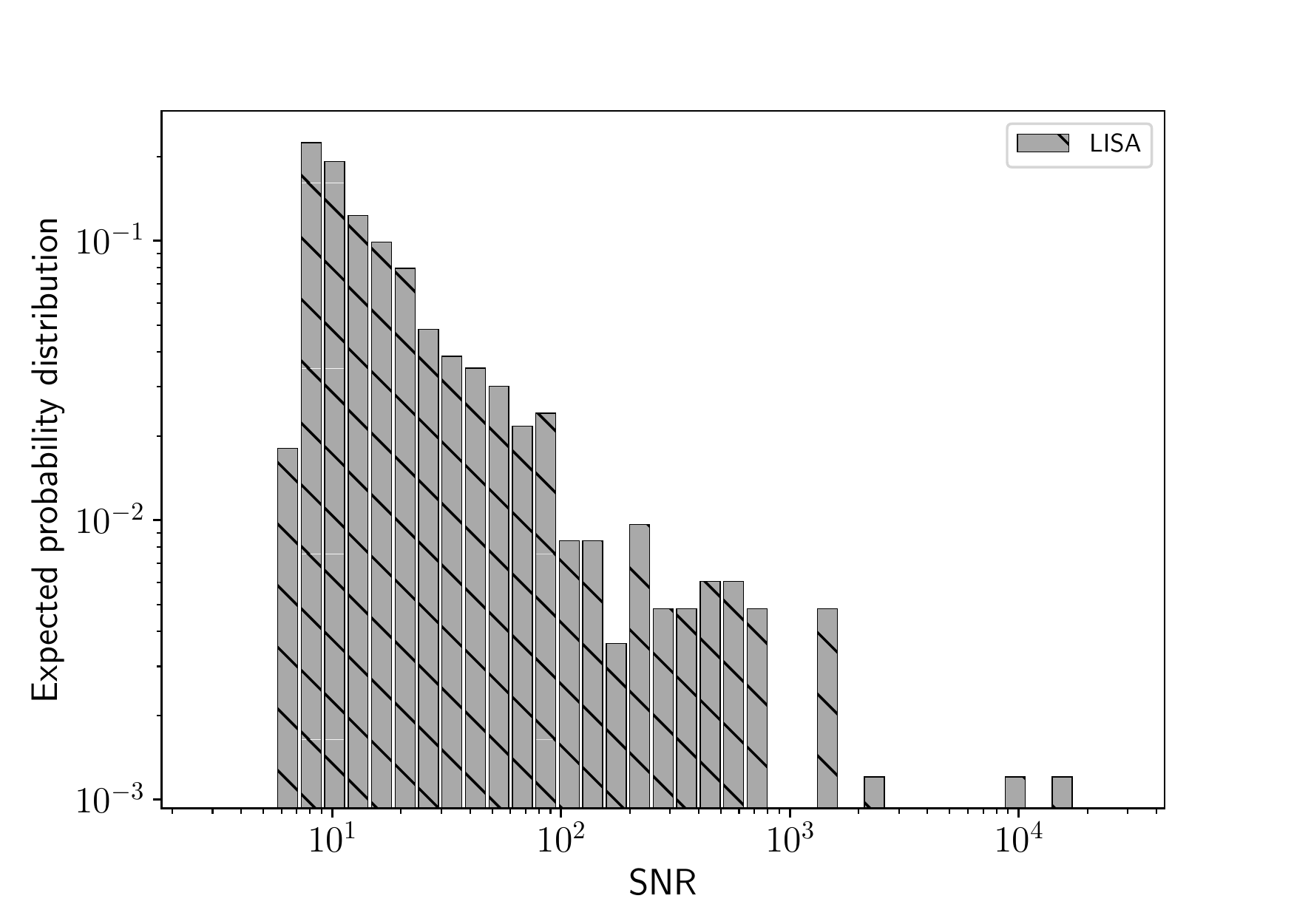}
        \label{fig:SNR_LISA_100MC}
    \end{minipage}
    % \hfill{}\hfill{}
    \begin{minipage}[l]{0.99\columnwidth}
        \centering
        \vspace{-.5cm}
        \includegraphics[scale=0.51]{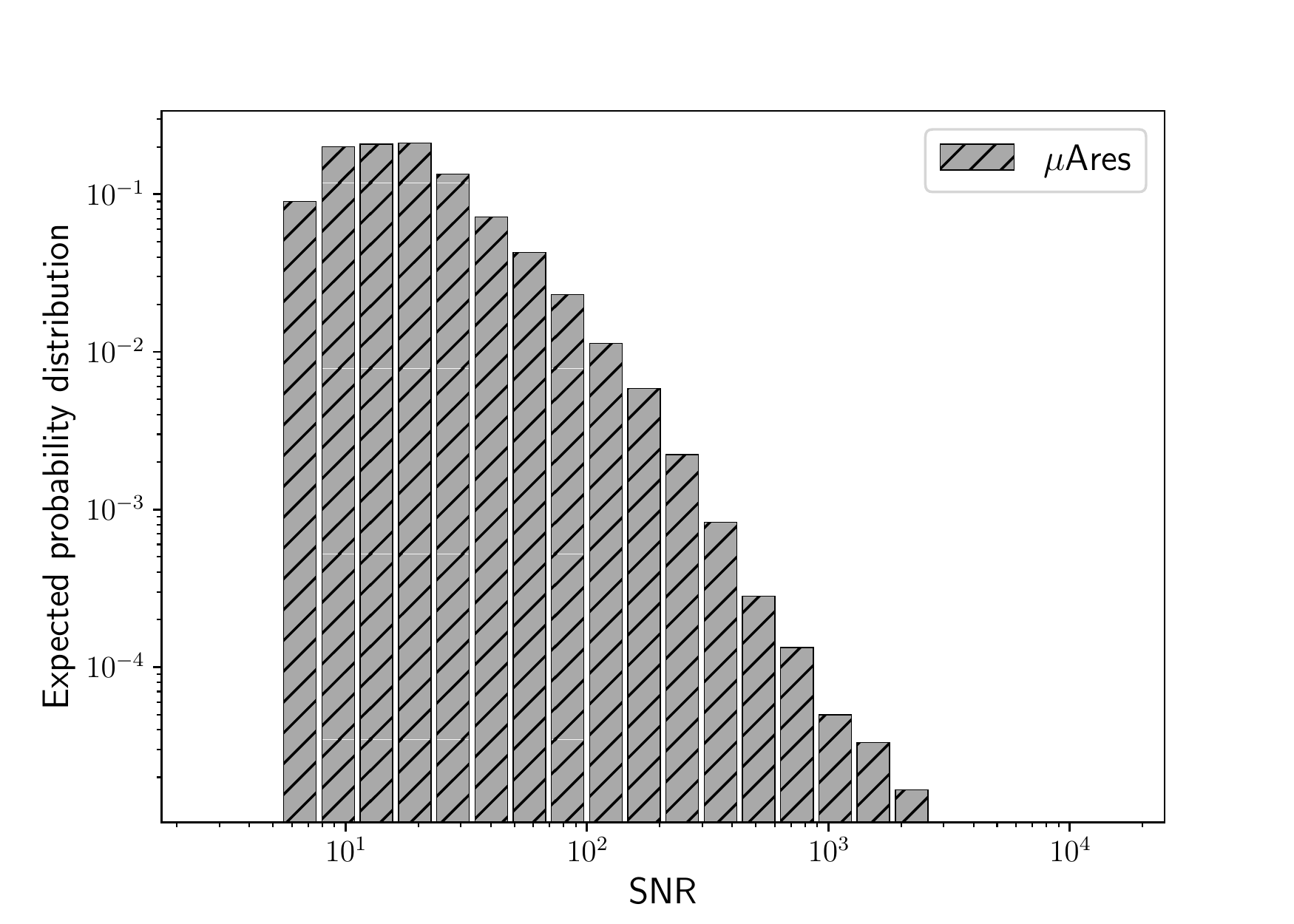}
        \label{fig:SNR_ARES_100MC}
    \end{minipage}
    \vspace*{-4mm}
    \caption{\label{fig:res}Resolvable sources' expected number (top row) and SNR (bottom row) probability distributions for LISA (left column) and $\mu$Ares (right column) from the entire PBH population. In the case of LISA this corresponds exactly to the same distributions from just the PBHs in the GW domain, as no resolvable PBHs are found inside the two body relax domain. For $\mu$Ares the $\approx300$ PBHs in the GW domain are always resolvable, while the remaining $\approx1000$ are found inside the two body relax region. Similarly, the contribution to the SNR distribution for $\mu$Ares is dominated by the more numerous fraction of PBHs in the two body relax region, whose SNR is generally $\,\lesssim30$.}
\end{figure*}

Here, $d$ is the luminosity distance from the source (8.26\,kpc \cite{gravity19} in the case of \sgrA), and $\dot{E_n}$ is the instantaneous power for a binary of chirp mass $\mathcal{M}_{c}=\mu^{3/5}m^{2/5}$ %\as{Define chirp mass. Also in a footnote is fine.}
at orbital frequency $f_{\mathrm{orb}}$, 
\begin{equation}
\label{eq:dotE}
    \dot{E_n}=\frac{32}{5}\frac{G^{7/3}}{c^5}(2\pi f_{\mathrm{orb}}\mathcal{M}_{c})^{10/3}g(n,e).
\end{equation}
The time frequency shift is given by 
\begin{equation}
    \dot{f} = \frac{96}{5}(2\pi)^{8/3}\Bigl(\frac{G \mathcal{M}_{c}}{c^3}\Bigr)^{5/3}f^{11/3}\times\mathcal{F}(e),
    \label{eq:fdot}
\end{equation}
where the term $\mathcal{F}(e)$ is
\begin{equation}
    \mathcal{F}(e)=\frac{1}{(1-e^2)^{7/2}}\Bigl(1+\frac{73}{24}e^2+\frac{37}{96}e^4 \Bigr),
\end{equation}
and the function $g(n,e)$ in Eq.\,(\ref{eq:dotE}), expressing the fraction of GW power going into each $n$-th harmonic, is \cite{petersmathews63}:
\begin{equation}
\label{eq:gne}
\begin{split}
&g(n,e)=\frac{n^4}{32}[J_{n-2}(ne)-2eJ_{n-1}(ne)+\frac{2}{n}J_{n}(ne)\\&+2eJ_{n+1}(ne)
-J_{n+2}(ne)]^2+(1-e^2)[J_{n-2}(ne)\\
&-2J_{n}(ne)+J_{n+2}(ne)]^2+\frac{4}{3n^2}[J_{n}(ne)]^2. 
\end{split}
\end{equation}
Here $J_n$ represent the n-th order Bessel functions of the first kind. 

%%%%%%%%%%%%%%%%%%%%%%%%%%%%%%%%%%
\section{\label{sec_eccresults}Results}

%%%%%%%%%%%%%%%%%%%%%%%%

\begin{figure*}[!ht]
    \begin{minipage}[l]{.91\columnwidth}
        \centering
        \includegraphics[scale=0.41]{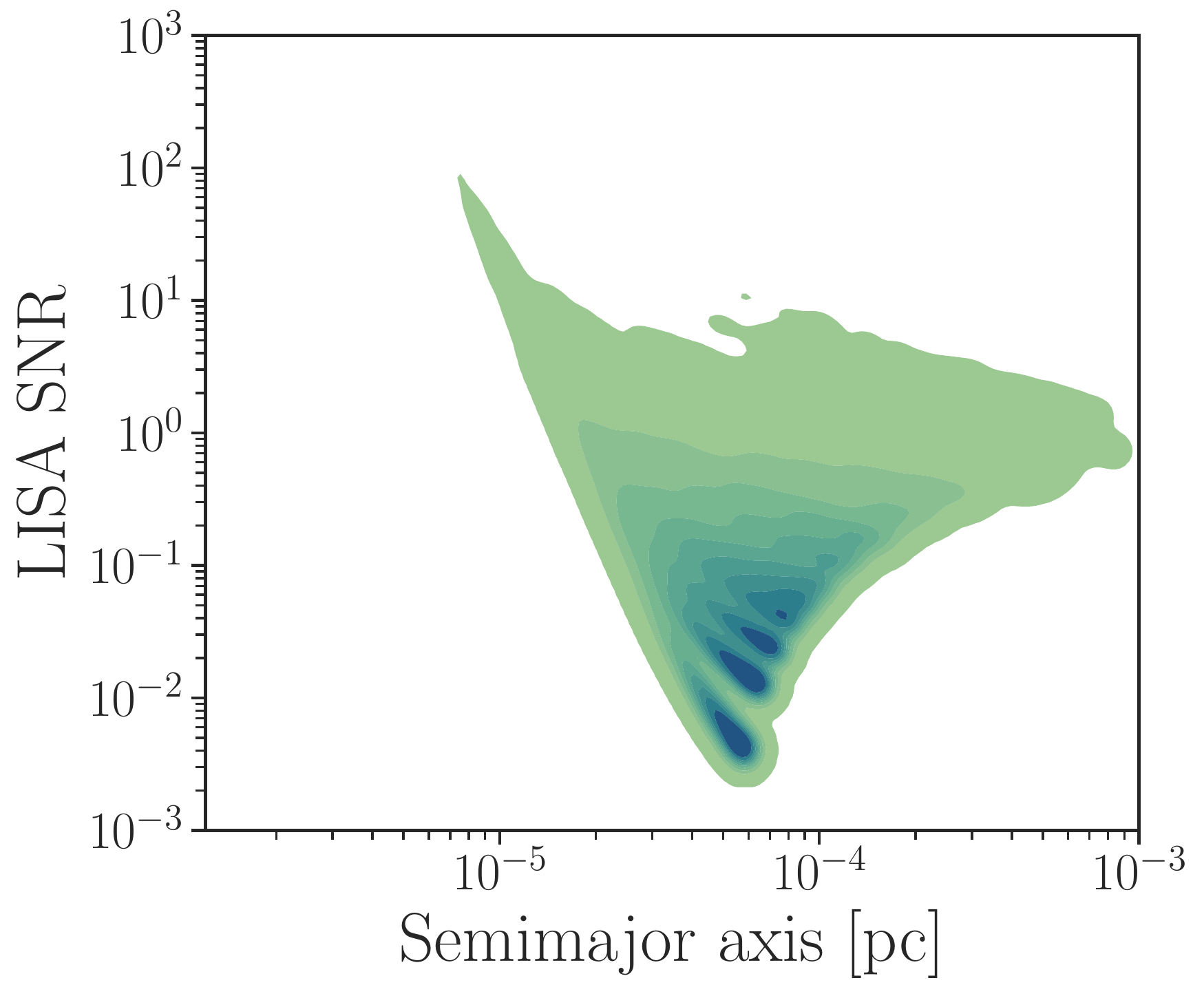}
        \label{fig:contsL}
    \end{minipage}
    % \hfill{}\hfill{}
    \begin{minipage}[l]{0.91\columnwidth}
        \centering
        \includegraphics[scale=0.41]{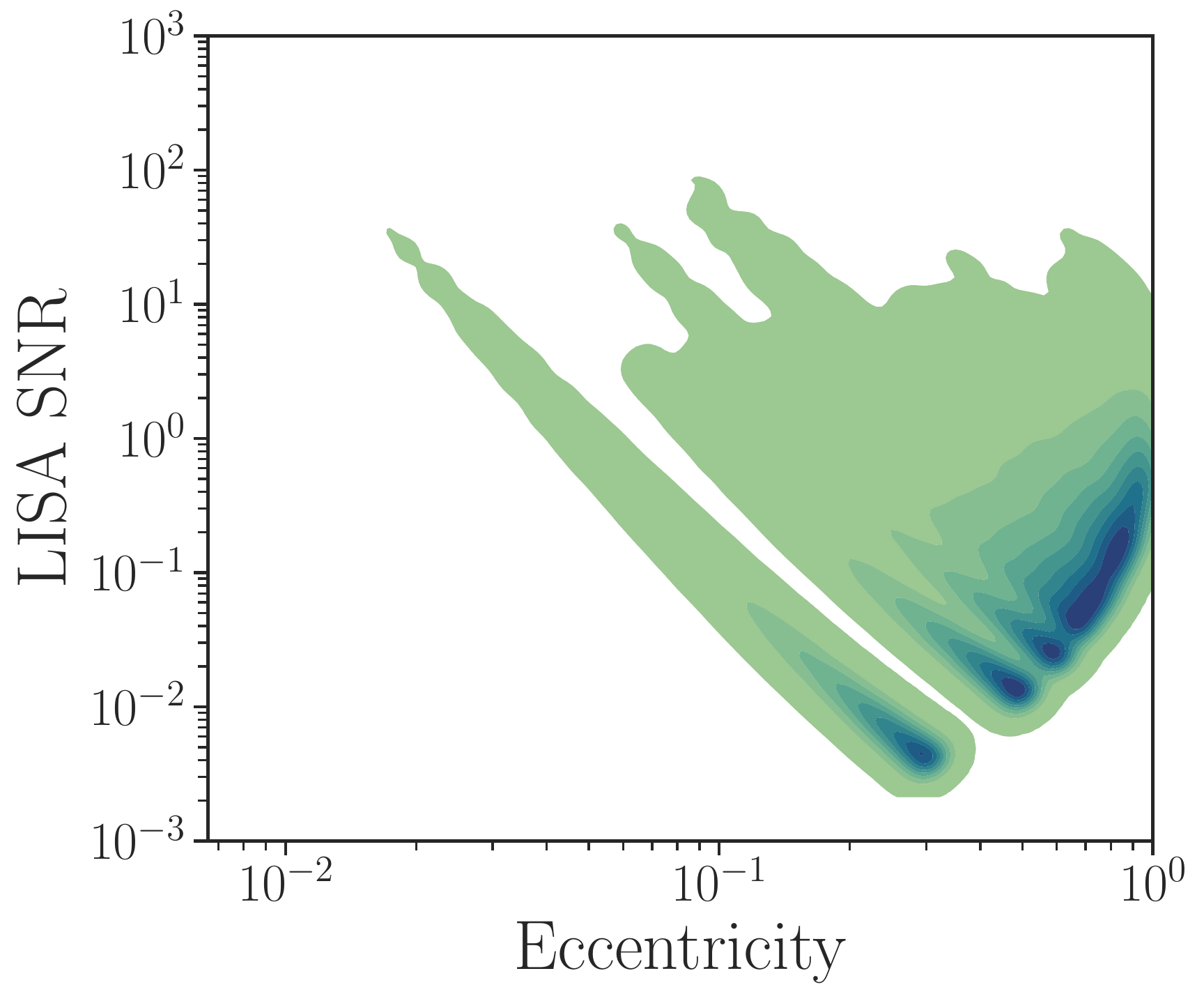}
        \label{fig:conteL}
    \end{minipage}\\
    % \hfill{}
    \begin{minipage}[l]{0.91\columnwidth}
        \centering
        \includegraphics[scale=0.41]{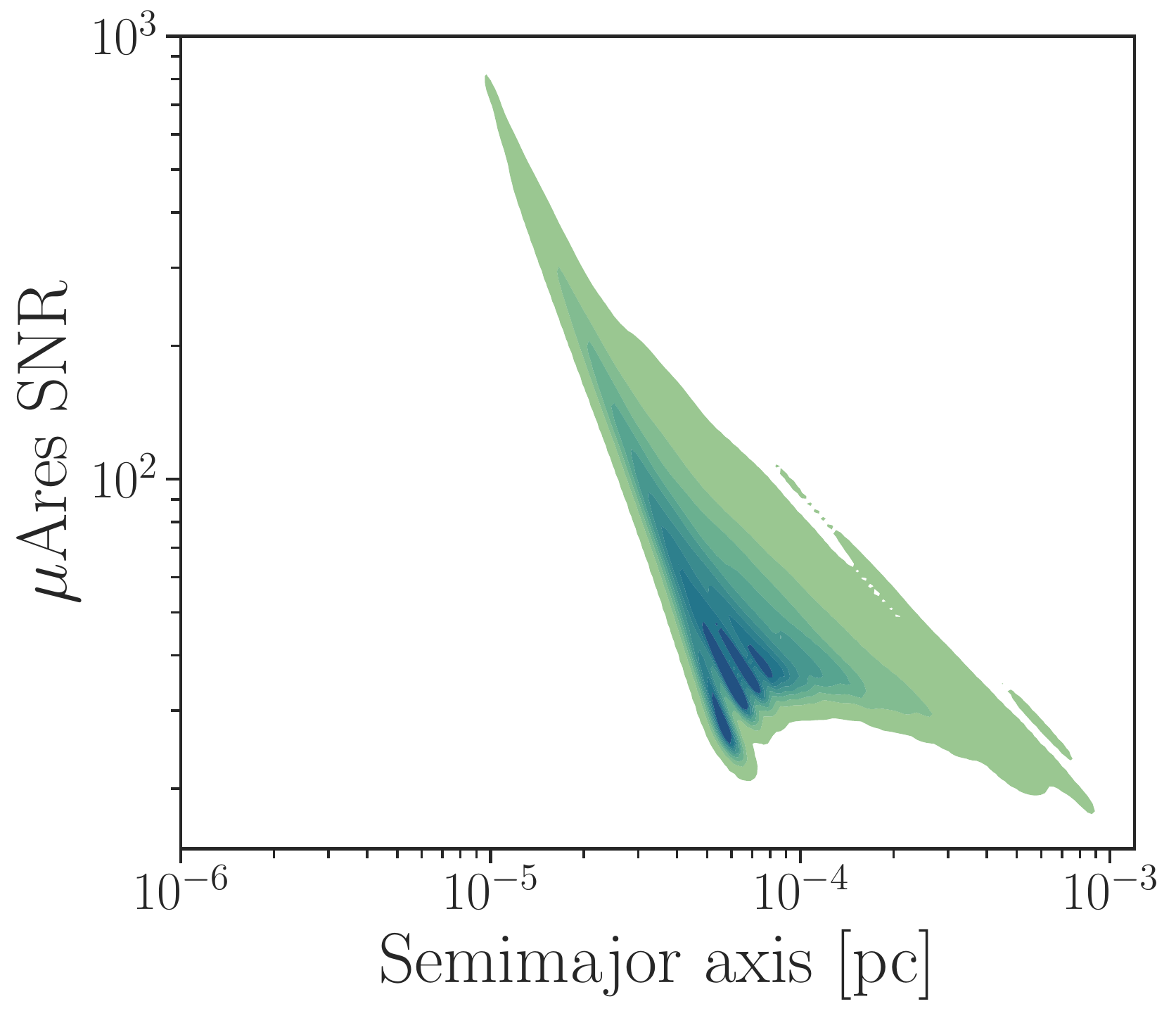}
        \label{fig:contsa}
    \end{minipage}
    % \hfill{}\hfill{}
    \begin{minipage}[l]{0.91\columnwidth}
        \centering
        \includegraphics[scale=0.41]{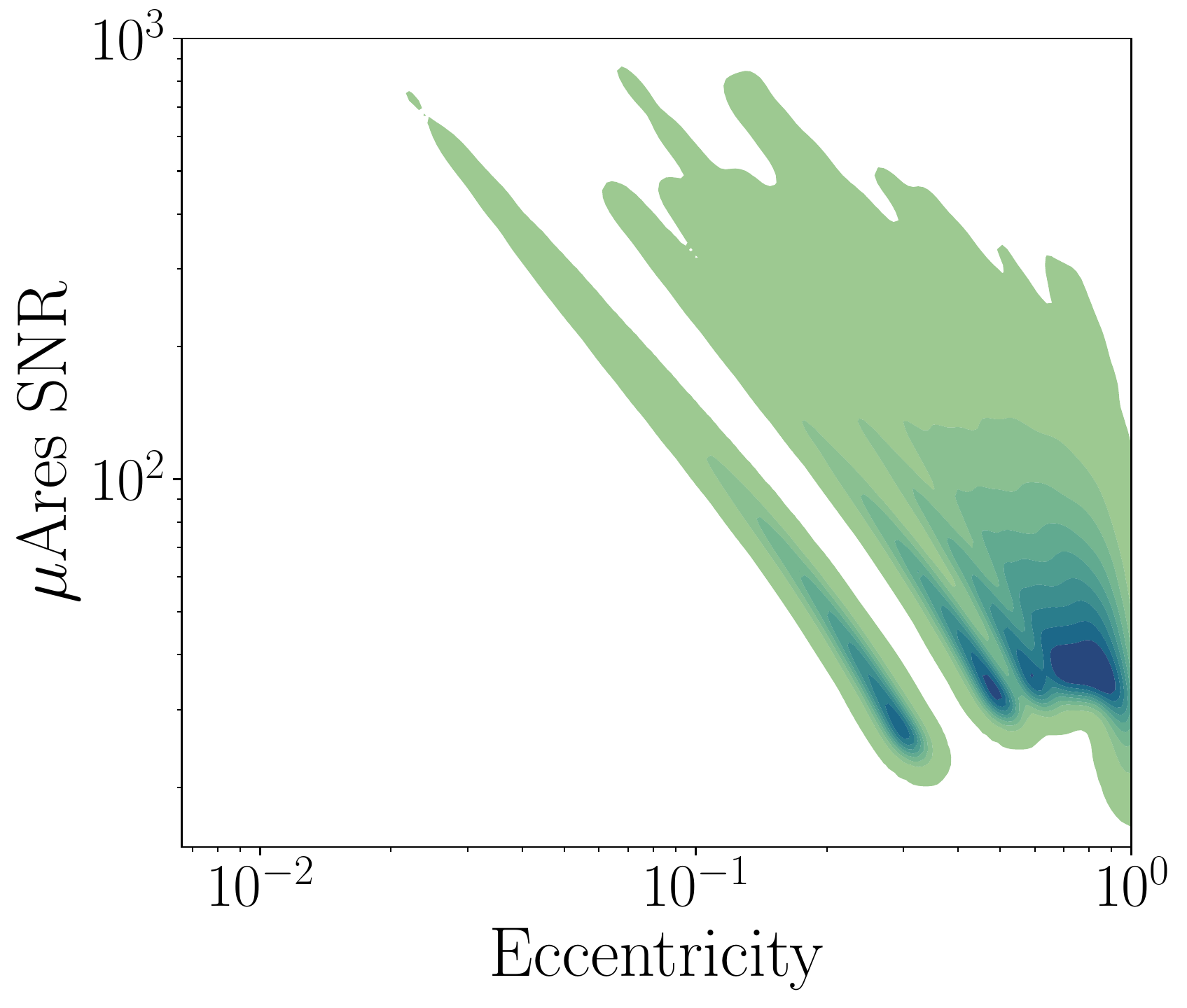}
        \label{fig:conteA}
    \end{minipage}
    % \vspace*{-1mm}
    \caption{\label{fig:cont}Contour plots showing the SNR vs semi-major axis (left column) and SNR vs eccentricity (right column) for LISA (top row) and $\mu$Ares (bottom row) density estimations from 1,000 Monte Carlo realizations referred to all the PBHs in the GW domain. The levels correspond to iso-proportions of the density, with darker layers indicating higher estimations. The cutoff for the lowest probability layer is at the 1\textperthousand~probability for every panel (i.e.~events with lower probability are not represented).\comm{; the color gradient refers to probability density function in $\textrm{d}\log$.}}
\end{figure*}

Following the procedure explained in Section\,\ref{sec_eccdensity} and \ref{sec_eccGW} we proceed now to evaluate the resulting GW signal from the entire PBH population, assessing its properties and detectability. 
Because PBHs that are still in the two body relaxation domain do emit gravitational radiation as much as those in the GW domain, we will obtain our results for the combined shares of the whole PBH population, which on average is split in sub-populations of 300 and 3700 PBHs between the two domains. We will further comment where appropriate on the relative contribution to the total results from the two sub-groups. We run a set of 1,000 Monte Carlo realizations of the population of PBHs, and average the results accordingly.\\
Due to GW emission, on average the innermost PBHs show a small residual eccentricity, resulting in a somewhat larger strain because of their proximity to \sgrA. On the contrary, the outermost PBHs tend to retain their generally higher eccentricity, while their greater distance to \sgrA~result in lower strain amplitudes.\\ 
For each PBH we consider the first 1,000 harmonics, implying that we may have some contribution to the observed GW strain from frequencies as high as $10^{-1}$\,Hz, since $f_n=f_{\txt{orb}}\times n$ and the highest orbital frequencies are just shy of $\sim 10^{-4}$\,Hz. As an example, in Fig.\,\ref{fig:hcharm} we show the GW strain from the harmonics up to $n=1000$, each represented with a different color, generated by an hypothetical PBH with initial eccentricity $e_0=0.92$ and orbital frequency $f_0=7\times10^{-6}$\,Hz.

Conditions for a significant frequency shift during observations, i.e., high eccentricity and small semi-major axis, do not coexist simultaneously for the PBHs in our model. Indeed, the frequency shifts over 10 years are in general negligible. As an example, we plot in Fig.\,\ref{fig:cfr_evo} the evolution in frequency, eccentricity and semi-major axis for three binaries of different initial eccentricity $e_0$ ($0.3$, $0.67$ and $0.92$), representative of a low, average, and high initial eccentricity among the distribution, over a time span of 10 years.
The initial semi-major axes (or equivalently, the initial frequencies) are chosen in agreement with a standard generic population of PBHs as depicted in Fig.\,\ref{fig:popolazione}, where the three are represented as red stars. In all three cases, the evolution of orbital parameters over 10 years is in fact completely negligible. In other words, the population of PBHs is by all means stationary in the $[a, e]$ space, implying  a characteristic strain in the form of a series of discrete delta functions (or very narrow lines), one for each harmonic considered. This fact could have been inferred as well from Fig.\,\ref{fig:hcharm}, considering that each colored segment, representative of the evolution in frequency of the strain over 10 years, is in fact reduced to the size of the markers for every harmonic.

\begin{figure*}
    \begin{minipage}[l]{.99\columnwidth}
        \centering
        \includegraphics[scale=0.45]{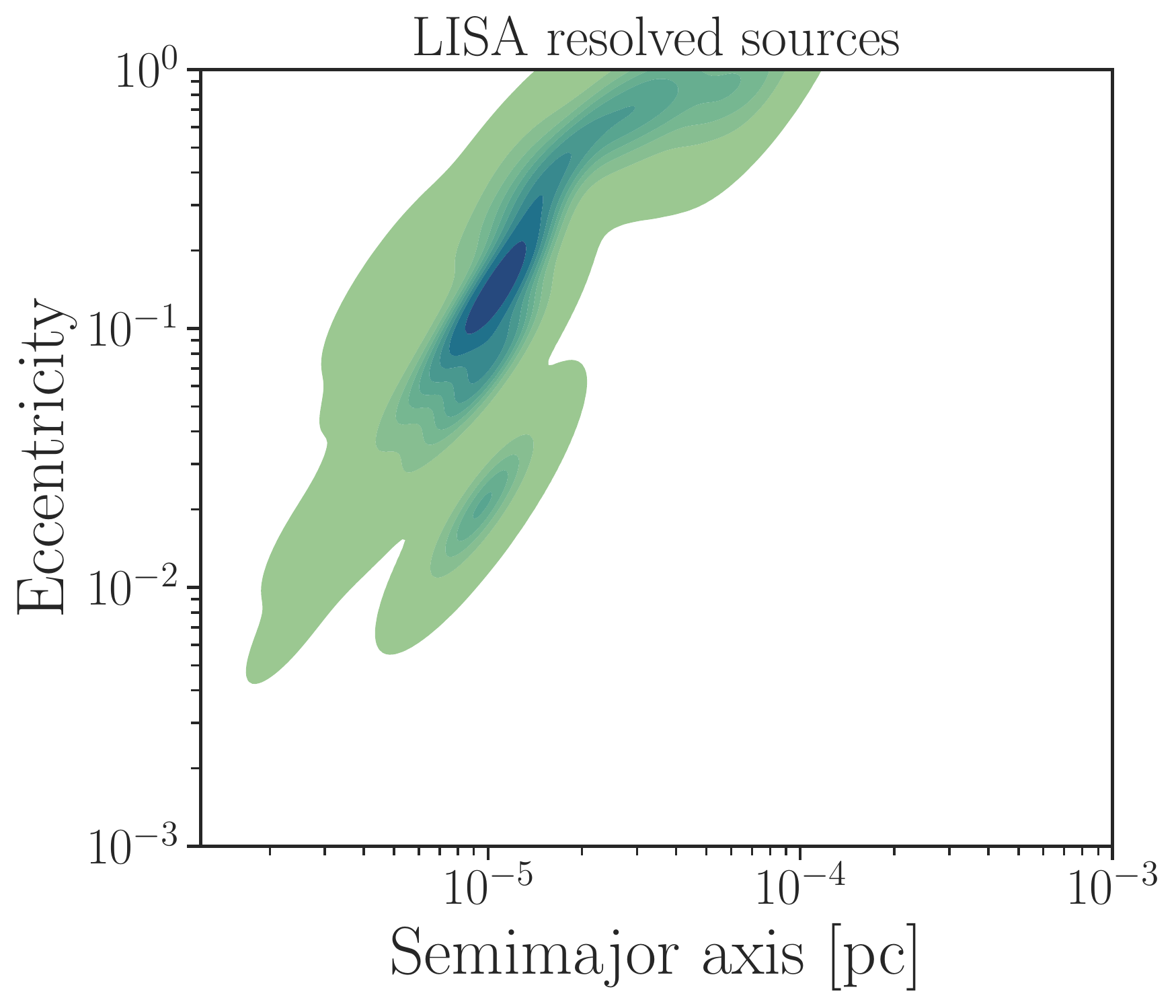}
        \label{fig:contseL}
    \end{minipage}
    % \hfill{}\hfill{}
    \begin{minipage}[l]{0.99\columnwidth}
        \centering
        \includegraphics[scale=0.45]{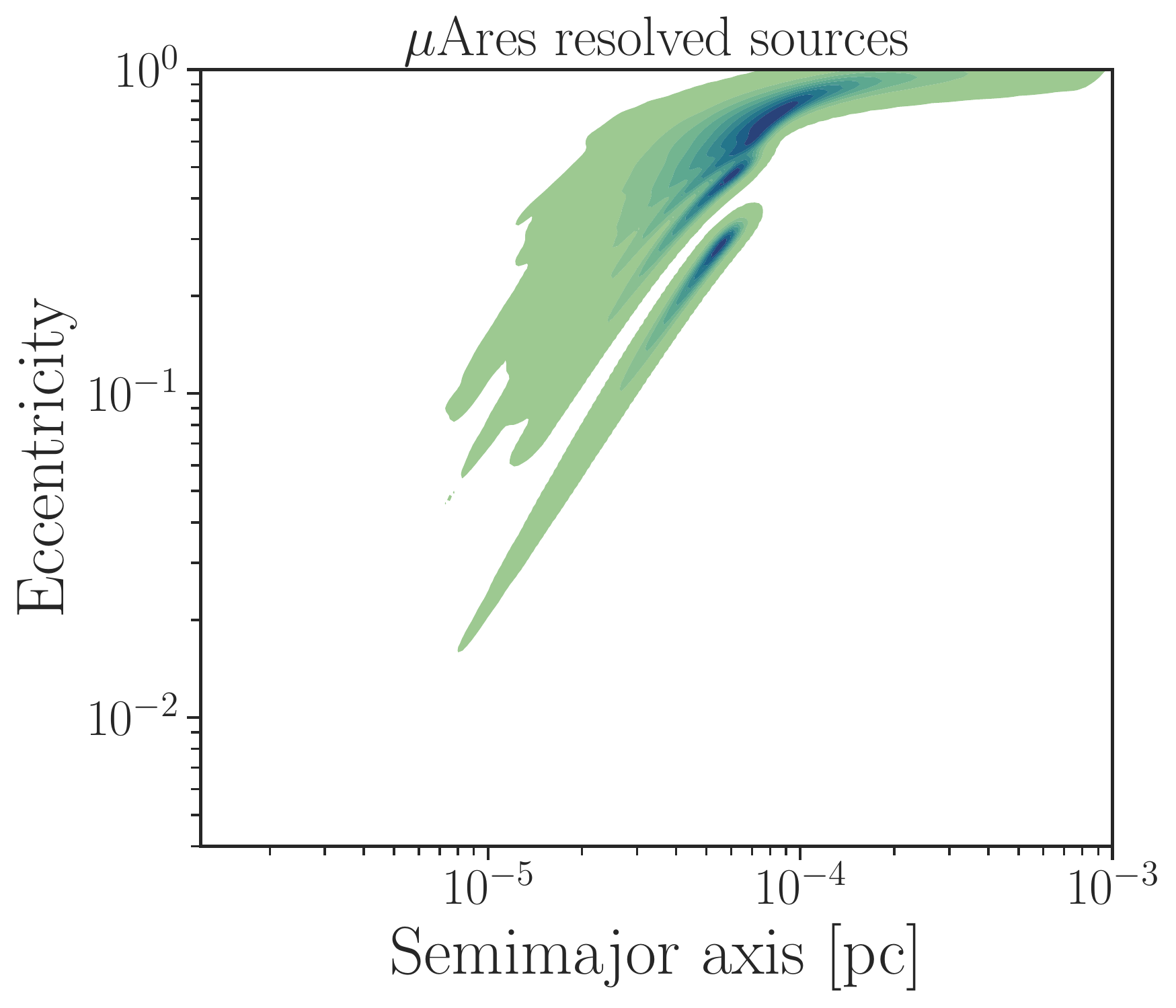}
        \label{fig:contseA}
    \end{minipage} \caption{\label{fig:cont_res}Contour plots showing the eccentricity vs semi-major axis density estimations from 1,000 Monte Carlo realizations of all the sources in the GW domain for LISA (left) and $\mu$Ares (right). The cutoff at the lowest layer is at the 1\textperthousand~probability for both.}
\end{figure*}
\subsection{Resolved sources}

We first assess the GW signal due to PBHs that can be treated as resolved sources. Following \cite{2010PhRvD..81f3008B, sesa08, bondani}, we deem a source as ``resolved" (or resolvable) when its signal-to-noise ratio SNR (sometimes indicated as $S/N$) is larger than 8. The SNR is generally computed as% \vspace*{-.5cm}
\begin{equation}
% \vspace*{-.5cm}
    \textrm{SNR}=\sqrt{\int\frac{h_{c}^2(f)}{f^2S_{\textrm{noise}}(f)}\mathrm{d}f},
    \label{eq:snr}
\end{equation}
where $h_c$ is the characteristic strain of the source as a function of frequency $f$, and $S_{\textrm{noise}}(f)$ is the interferometer's sky averaged power spectral density. Since our PBHs do not evolve in frequency appreciably during an estimated mission lifetime of 10 years (see again Fig.\,\ref{fig:cfr_evo}), the integral in equation~\eqref{eq:snr} is a sum over the signal harmonics. An example of Monte Carlo realization of the PBH population is shown in Fig.\,\ref{fig:res_SNR}. This particular realization counts 294 PBHs in the GW-driven regime. Specifically, we show, as a function of semi-major axis, the distribution in eccentricity (green triangles), SNR referred to the LISA sensitivity (black diamonds) and an estimate of the harmonic carrying the largest strain (purple stars).
The harmonic corresponding to the maximum GW emitted power can be estimated as a function of the PBH initial eccentricity $e_0$ as (see e.g. \cite{2003ApJ...598..419W})
\begin{equation}
    n_{\txt{max}} = 2\times\frac{(1+e_0)^{1.1954} }{ (1-e_0^2)^{1.5}}.
    \label{eq:n_max}
\end{equation}
\noindent The top left and top right panels of Fig.\,\ref{fig:res} show the probability distributions of the number of sources resolved by LISA and $\mu$Ares, respectively, while the bottom left and bottom right panels show the corresponding distributions of SNRs. All the resolvable sources for LISA are found in the GW domain, while generally $\mu$Ares will resolve all the $\approx300$ PBHs in the GW domain plus, on average, $\approx1000$ from the two body relaxation domain.\\
Comparing Fig.\,\ref{fig:res} with the same results from \cite{bondani}, who assumed circular orbits, the probability of resolving at least one PBH with LISA over 10 years of observation increases from $10\%$ to $60\%$, in virtue of the introduction of an eccentricity profile. Note that there is also a non negligible 20\% chance of detecting more than one PBH in the same conditions.\\
The situation of $\mu$Ares is completely different, as the expected number distribution of resolvable sources appears to closely match the total number of PBHs in the GW domain of Fig.\,\ref{fig:popolazione}. This is not surprising, considering the much better sensitivity of $\mu$Ares in the $\mu$Hz band compared to LISA. \\
The probability distribution of the expected SNRs for LISA (Fig.\,\ref{fig:res}, bottom left panel) appears to peak at the detection threshold of SNR $=8$ with an extended tail at larger SNRs. The sparse detections with SNR $>1000$, generally arise from PBHs that happen to be at an advanced stage in their evolution track (i.e.~closer to \sgrA, see Fig.\,\ref{fig:popolazione}), rather than from PBHs initially particularly highly eccentric. In general a binary with $e>0.99$ will evolve very quickly in comparison with less eccentric binaries, eventually merging after a comparatively much shorter time, making it less likely to be caught by LISA over its 10 years observation time. 
The PBHs in our model are by the same means approximately persistent, as discussed in regards of Fig.\,\ref{fig:cfr_evo}. \\
Comparing again our results with the circular orbit scenario of \cite{bondani}, we found a clear increase in the number of detectable sources, while the corresponding SNR distribution is approximately similar.\\
With $\mu$Ares being capable of resolving all the PBHs in the GW regime, and almost 30\% of the PBHs in the two body relax regime, the probability distribution in the bottom right panel of Fig.\,\ref{fig:res} covers more smoothly the range to the higher end of the SNR spectrum. 
\begin{figure*}[!ht]
    \begin{minipage}[l]{\columnwidth}
        \centering
        \includegraphics[scale=0.52]{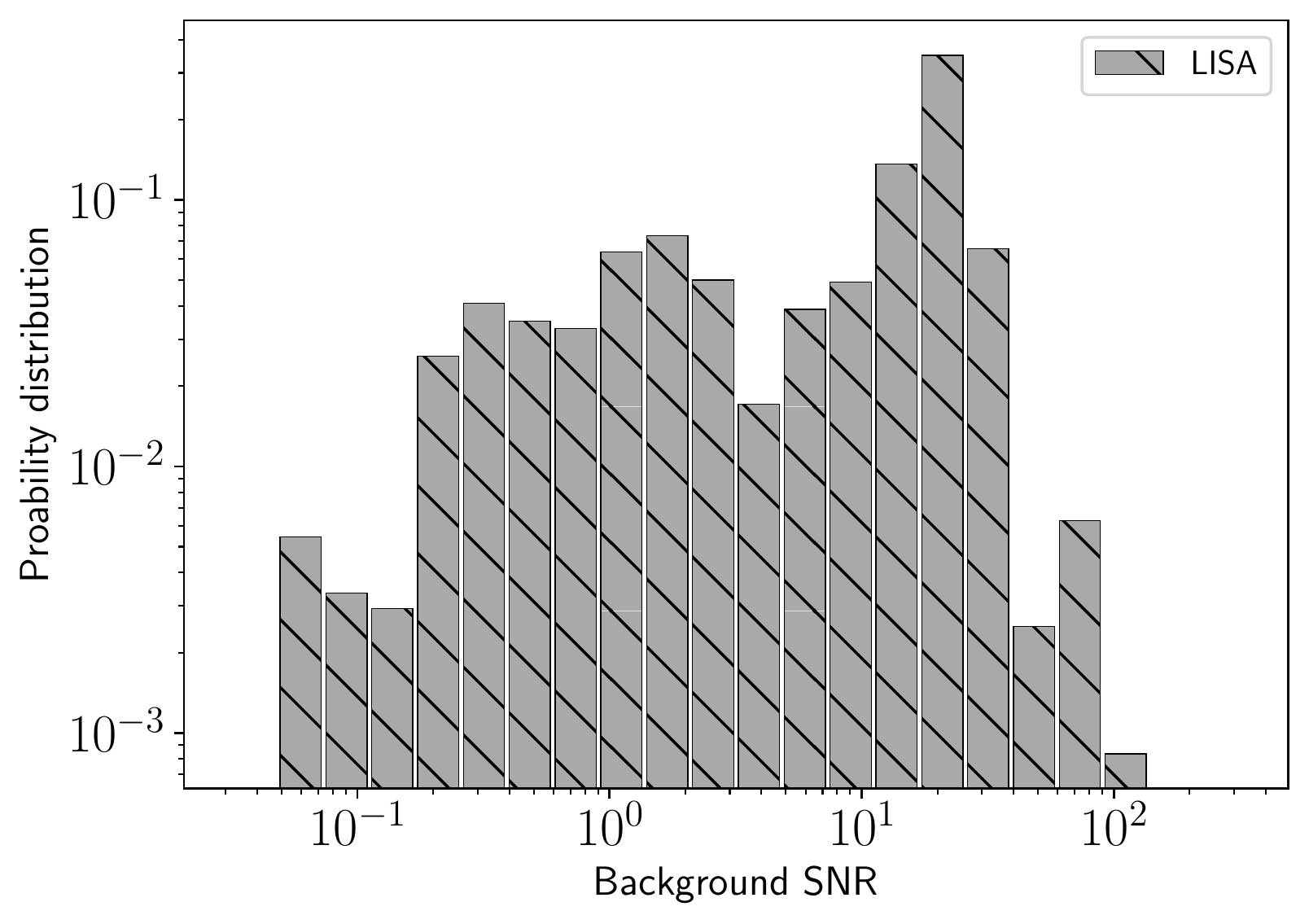}
        \label{fig:BG_L_tot}
    \end{minipage}
    \begin{minipage}[l]{\columnwidth}
        \centering
        \includegraphics[scale=0.52]{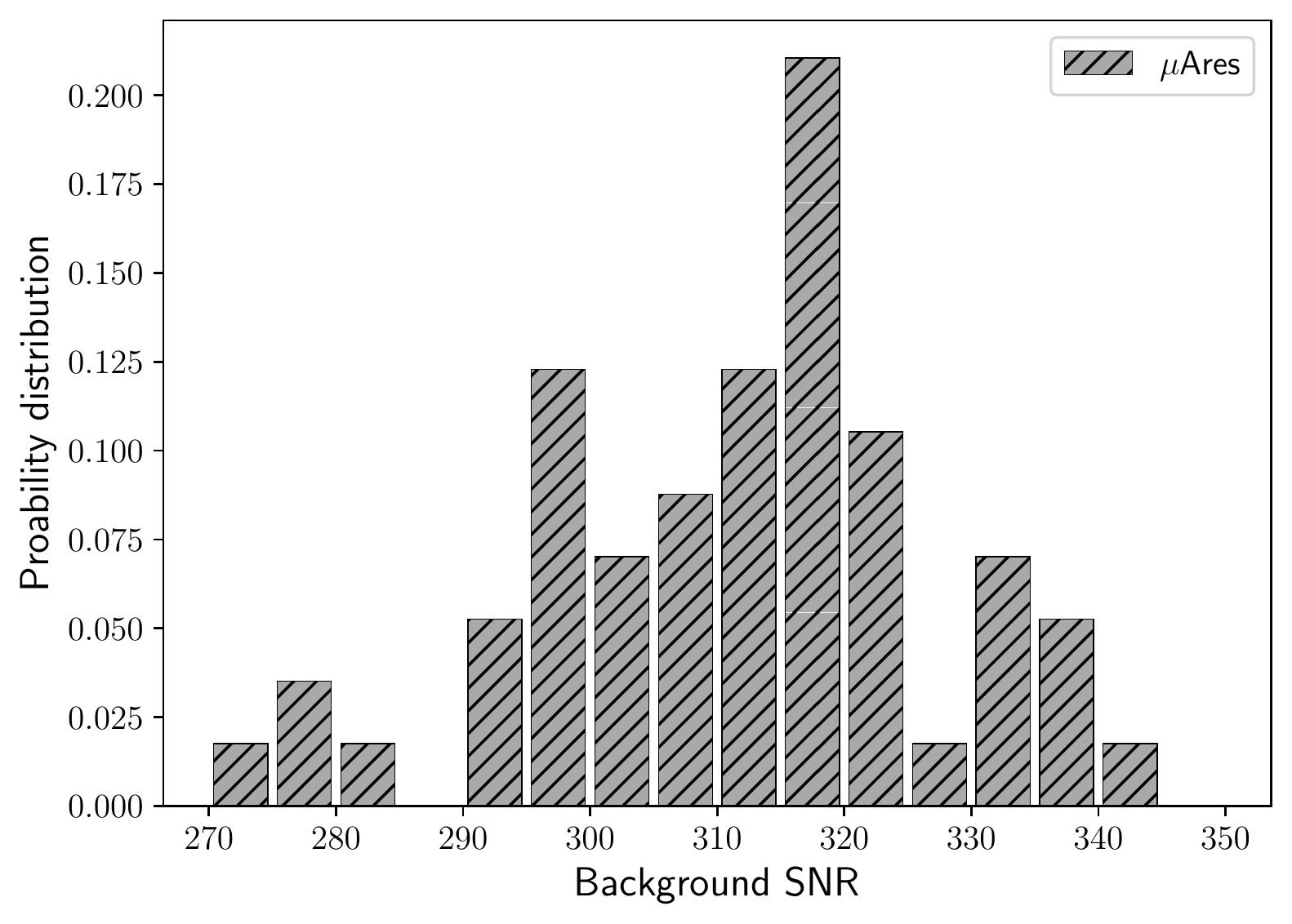}
        \label{fig:SNR_ARES_BG_ext}
    \end{minipage}
    % \vspace*{-1mm}
    \caption{\label{fig:BG_tot} GW background for LISA (left) and $\mu$Ares (right) from the entire PBH population. In the case of LISA the distribution has two modes, the first centered around SNR $\simeq4$ determined by the PBHs in the GW domain, and the second centered around SNR $\simeq20$ determined by the PBHs in the two body relax domain. In the case of $\mu$Ares the background is determined solely by the unresolved fraction of the PBH population in the two body relaxation regime.}. 
\end{figure*}
In Fig.\,\ref{fig:cont} we show contour plots (displayed probability $>1$\textperthousand) of SNR vs semi-major axis (left column) and SNR vs eccentricity (right column) referred to the PBHs in the GW domain for the complete set of 1,000 Monte Carlo realizations: we can see for LISA on the top row and for $\mu$Ares on the bottom row the contour extending to high SNRs at low semi-major axis, where the eccentricity is generally small, and compare with the same data plotted against eccentricity on the right column. The two orbital parameters, $a$ and $e$, for the resolved PBHs in the GW domain, are themselves distributed as shown in the contour plots in Fig.\,\ref{fig:cont_res}.\footnote{We must note that the non continuous behavior of the probability maps showed in Fig.\,\ref{fig:cont} and Fig.\,\ref{fig:cont_res} is an artifact due to the numerical discretization we adopted in the initial values of the semi-major axis. Such discretization is apparent in  Fig.\,\ref{fig:popolazione} and Fig.\,\ref{fig:res_SNR}. As an example, the lower isolated ``island" appearing in the right panel of Fig.\,\ref{fig:cont_res} reflects the uppermost evolutionary track in Fig.\,\ref{fig:popolazione}.}

\subsection{GW Background}
The cumulative signal from the non resolvable PBHs adds up to the GW background. We will compute here the resulting SNR from the entire PBH population.
The amplitude of the GW background signal is computed using Eq.\,(7) of \cite{sesa16},
%%%%%%%%%%%%%%%%%%%%%%%%%%%%%%%%%%%%
\begin{equation}
(S/N)^{2}_{\txt{bkg}} = t_{\txt{obs}}\int\gamma(f)\frac{h_{c,\txt{bkg}}^{4}(f)}{4f^{2}S_n(f)^{2}} df,
\label{eq:SNRbkg}
\end{equation} 
where the response function in the relevant frequency range is $\gamma(f)\approx1$ according to Fig.\,4 of \cite{thraneromano} and to \cite{sesa16}, while $h_{c,\txt{bkg}}^2$ is intended as the characteristic strain of the background, resulting from the non-resolvable sources, summed over each frequency bin $\Delta f=t_{\txt{obs}}^{-1}$. In particular, following the method in \cite{bonetti} for the computation of the background, given the $n$-th harmonic's strain rms, defined in \cite{finn, triplets} as:
\begin{equation}
    h_n^2=\frac{G\dot{E}_n}{c^3\pi^2d^2f_n^2}
\end{equation}
the characteristic strain for the GW background of Eq.\,(31) in \cite{bonetti} takes for our model the simplified form:
\begin{equation}
   h^{2}_{c,\txt{bkg}}(f) = \int \mathrm{d}e~\times~ \Bigr[\sum_{n}\frac{\mathrm{d}^2N}{\mathrm{d}e\,\mathrm{d\,ln}f_{\txt{orb}}}h_n^2(f) \Bigl]
\end{equation}
It is important to point out that each frequency bin contains not only the signal from the second harmonic (i.e.~$2f_{\txt{orb}}$) of a single PBH as would be the case with circular orbits, but all the different harmonics generated from the rest of the PBH population that happen to not differ by more than a single $\Delta f$ between each other. For 10 years of observation time, $\Delta f=3\times10^{-9}$\,Hz.\\
Comparing once again with the results from \cite{bondani}, the expected GW background SNR distribution for LISA (Fig.\,\ref{fig:BG_tot}) is dominated by the share of PBHs in the two body relaxation domain, whose GW background signal appears to be detectable with SNR\,$\simeq20$ thanks to the highly eccentric and numerous PBHs in this region. This result is roughly three orders of magnitude larger with respect to the circular orbit case. A subdominant mode from the $\approx300$ PBHs in the GW domain is found around SNR$\,\simeq4$. 
Concerning $\mu$Ares, the results obtained in the previous Section, implying that it should be able to resolve all the GW-emitting PBHs inside the GW domain, entail that only the PBHs from the two body relax domain contribute to the background, with a resulting SNR distribution centered around $\simeq300$.
In general, these results are driven by the very high eccentricities of the more external orbits, combined with a tenfold increase in the number of PBHs when moving to the relaxation region. On this note, the main driver for a PBH to be resolvable when eccentricities are high and the number of contributing harmonics can accordingly vary greatly, is its SNR being above detection threshold; because of this, the rare instances where inside a frequency bin $\Delta f$, two or more PBHs are found, are hardly indicative of such PBHs to be flagged as non-resolvable, since indeed, the dominant contributing harmonics are generally at sufficiently different frequencies.
The results obtained for $\mu$Ares are in agreement with its much better sensitivity in the $\mu$Hz band --~where most of the PBHs are statistically expected to be found~-- when compared to LISA: every PBH close enough to \sgrA~to be evolving by GW emission should be resolved. When looking at the entire population, the number of detectable PBHs increases on average to a total of $\approx1300$. The remaining non resolved sources contribute to build up a detectable background signal with SNR $\simeq300$.
\subsection{Results for different PBH masses}
Until now we have focused our attention on 1\,M$_{\odot}$ PBHs, based on the results of recent theoretical arguments by \cite{prefmass} suggesting that, under the assumption of a scale-invariant amplitude of primordial curvature fluctuations, the resulting PBH mass spectrum should show a clear peak at $\simeq 1\,M_{\odot}$. Furthermore this assumption provides for a simple tool for comparison with previous results in the literature. Nevertheless, PBHs can in principle possess any mass at formation, and although we made no explicit assumptions on the dark matter fraction in PBHs, we did assume all the diffuse mass inside the orbit of S2 to be in PBHs. Since a monochromatic
distribution of 1\,M$_{\odot}$ PBHs is already ruled out as a viable
candidate for all the dark matter by several observational constraints, as shown for example in \cite{carr2021}, if PBHs are to constitute a significant fraction of dark matter at all, it's possible they would do so with an extended mass function. Such an approach would go beyond the scope of this paper, but it is nevertheless instructive to observe how our results would be modified by a change in the PBH mass, while still adopting monochromatic mass functions. In accordance with theoretical arguments from \cite{carr1, carr2021} (among others), the possibility of finding a significant dark matter fraction in PBHs, while still maintaining relevance for GW detection, is to be found in the range $0.1<\mpbh<10$\,M$_{\odot}$. A PBH mass too much outside of this interval is most likely to yield negligible results for our scope, either in terms of GW detectability for lower masses, or because of too stringent limits on the dark matter fraction in PBHs at $\mpbh>10$\,M$_{\odot}$ \cite{carr2021}. For this reason we obtained further results referred to these two additional PBH masses, with all other model parameters unvaried. Naturally, because the total diffuse mass enclosed by the orbit of S2 is to remain unchanged, the total number of PBHs in our model varies accordingly, from a minimum of 400 for $\mpbh=10$\,M$_{\odot}$ up to 4$\times10^4$ for $\mpbh=0.1$\,M$_{\odot}$.\\
In the first case of the two, $\simeq$ 30 PBHs are found on average in the GW domain. For LISA, the detectability of at least 1 PBH has a probability of 58\%, including a 21\% chance of at least 2 detections, with up to 5 PBHs resolved in a single Monte Carlo realization. The corresponding SNR distribution is skewed around the detectability threshold of 8 and we observed a 85\% probability of SNR $<$ 20.\\
For $\mu$Ares, $\simeq$ 250 PBHs are likely resolvable, with average SNR $\simeq$ 300. The background GW signals have distributions centered around SNR = 80 for LISA (with 75\% probability of background SNR $>8$), and centered around SNR = 350 for $\mu$Ares.\\
In the second case, i.e., $\mpbh=0.1$\,M$_{\odot}$, on average $\simeq3000$ PBHs are found inside the GW domain. Because of their low mass, LISA has a negligible ($<1\%$) chance of resolving any PBHs among the entire population, while the GW background signal is by all means undetectable, the SNR never exceeding 0.5. Once again the situation is different for $\mu$Ares, which is able to resolve $\simeq 900$ PBHs, all found inside the GW domain (PBHs in the two body relax region only have SNRs as high as 3.5). By contrast the GW background is detectable thanks to the numerous share of the PBH population in the two body relax region, building up a GW background signal with SNR~$\simeq180$.
% internal BG (10\% probability) could be observed with $3<$SNR $<6$

%%%%%%%%%%%%%%%%%%%%%%%%%%%%%%%%%%%%%%%%%%%%%%%%%
\section{\label{sec_concl}discussion and conclusions}

The detection of PBHs with GW observations from the GC with LISA and $\mu$Ares might prove crucial in helping to solve the dark matter problem. In this work we expanded from the recent results in the literature by estimating the GW signal from a 1\,$M_{\odot}$ PBH population orbiting \sgrA~characterized by a thermal distribution in eccentricity, and in compliance with the limits in total diffuse mass inside the orbit of S2 from the current best observational constraints. 

After having quantified the fraction of PBHs whose evolution is driven by gravitational radiation, as opposed to two body relaxation, we compared with the results obtained for circular orbits, finding the chances of detecting one PBH with LISA over a 10 years long mission lifetime, having increased from 10\% to 60\%; furthermore, there is a 20\% probability of resolving more than one PBH over the same timescale. By considering an eccentrically orbiting PBH population, GW emission is triggered at various higher harmonics of the orbital motion. Such contribution to the GW signal is even more noticeable when estimating the expected GW background, where the signal limited to the non resolved sources appears to build up a total signal detectable by LISA with SNR $\simeq20$, in contrast with the circular case where such probability was shown to be low.

Comparing with our previous results obtained for circular orbits, the presence of an eccentricity profile --~which allows more PBHs to enter the LISA band~-- proved to be more influential, rather than the SNR distribution, on the total number of detections. Because PBHs of very high eccentricity evolve much faster than those on more circular orbits, the timescales of orbital evolution and of instrumental observation (which we, rather optimistically, maxed out at 10 years) imply too small a chance of catching a PBH with both an anomalously high eccentricity \textit{and} small semi-major axis. The average resulting PBH population is therefore statistically made of quasi-stationary sources.\\
Finally, we allowed the PBH mass to vary in the next most favourable range for GW detection, i.e., $0.1\div10$\,M$_{\odot}$, the higher detectability resulting from the higher mass of the two.\\
In conclusion, and reiterating the arguments of \cite{bondani}, we underline that the next generation of ground-based interferometers such as the Einstein Telescope \cite{Punturo2010}, thanks to their sensitivity at higher frequencies compared to LISA and $\mu$Ares, will play a complementary role in the search for GWs emitted by PBHs, e.g., in the detection of binaries of such objects \cite{PBHB1, PBHB2}. Indeed, the prospects of genuine multi-frequency GW observations \cite{sesa16} should greatly increase our chances of testing the existence of such an elusive population of black holes in the forthcoming decades.

\acknowledgements
We thank Enrico Barausse for his valuable contribution and suggestions. A.S. acknowledges financial support provided under the European Union’s H2020 ERC Consolidator Grant ``Binary Massive Black Hole Astrophysics'' (B Massive, Grant Agreement: 818691). F.H. acknowledges funding from MIUR under the grant PRIN 2017-MB8AEZ. 

% %\FloatBarrier

% \appendix
% \setcounter{secnumdepth}{0}
% \section{APPENDIX A}

\bibliography{biblio_ecc_pbh}

\end{document}